\documentclass[runningheads]{llncs}

\usepackage{eccv}

\usepackage{eccvabbrv}
\usepackage{multirow}
\usepackage{graphicx}
\usepackage{enumitem}
\usepackage{booktabs}
\usepackage{amsmath}
\DeclareMathOperator*{\argmax}{arg\,max}
\DeclareMathOperator*{\argmin}{arg\,min}
\usepackage[accsupp]{axessibility}  % Improves PDF readability for those with disabilities.

\usepackage{hyperref}

\usepackage{orcidlink}

\begin{document}

\title{Not Just Change the Labels, Learn the Features: Watermarking Deep Neural Networks with Multi-View Data} 

\titlerunning{MAT}

\author{Yuxuan Li\inst{1} \and
Sarthak Kumar Maharana\inst{2} \and
Yunhui Guo\inst{2}}

\authorrunning{Y.~Li et al.}

\institute{Harbin Institute of Technology, China  \and
The University of Texas at Dallas, Richardson, USA\\
\email{lyxzcx@outlook.com, \{skm200005, yunhui.guo\}@utdallas.edu}
}

\maketitle

\begin{abstract}
With the increasing prevalence of Machine Learning as a Service (MLaaS) platforms, there is a growing focus on deep neural network (DNN) watermarking techniques. These methods are used to facilitate the verification of ownership for a target DNN model to protect intellectual property. One of the most widely employed watermarking techniques involves embedding a trigger set into the source model. Unfortunately, existing methodologies based on trigger sets are still susceptible to functionality-stealing attacks, potentially enabling adversaries to steal the functionality of the source model without a reliable means of verifying ownership. In this paper, we first introduce a novel perspective on trigger set-based watermarking methods from a feature learning perspective. Specifically, we demonstrate that by selecting data exhibiting multiple features, also referred to as \emph{multi-view data}, it becomes feasible to effectively defend functionality stealing attacks. Based on this perspective, we introduce a novel watermarking technique based on Multi-view dATa, called MAT, for efficiently embedding watermarks within DNNs. This approach involves constructing a trigger set with multi-view data and incorporating a simple feature-based regularization method for training the source model. We validate our method across various benchmarks and demonstrate its efficacy in defending against model extraction attacks, surpassing relevant baselines by a significant margin. The code is available at: \href{https://github.com/liyuxuan-github/MAT}{https://github.com/liyuxuan-github/MAT}.
\end{abstract}

\section{Introduction}
\label{sec:intro}

\vspace{-0.3cm}

Deep neural networks (DNNs) have demonstrated superior performance across various tasks, including computer vision \cite{he2016deep,he2017mask,kirillov2023segment}, natural language processing \cite{devlin2018bert,lopez2017deep,kamath2019deep}, and speech recognition \cite{hannun2014deep,deng2014ensemble,zhang2018deep}. Their remarkable performance has led to widespread adoption in Machine Learning as a Service (MLaaS) platforms, allowing users to submit input data and retrieve model-generated outputs from the cloud \cite{ribeiro2015mlaas,yao2017complexity,kim2018nsml}.

MLaaS empowers users to leverage the strong capabilities of DNNs but also exposes MLaaS providers to potential risks. Specifically, as these providers invest significant resources in developing the source model, protecting their intellectual property rights becomes a critical concern. Although the attackers do not have access to the source model nor the source training data, the attackers can still extract the DNN models' functionality using black-box functionality stealing attacks \cite{orekondy2019knockoff,ma2021simulating,papernot2017practical,wang2022black}. For example, a model extraction attack leverages a surrogate model for imitating the output of the source model on a surrogate dataset for functionality stealing \cite{orekondy2019knockoff}. 

To enable the ownership verification of a stolen model, one of the most effective approaches is to use a trigger set for model watermarking \cite{adi2018turning,zhang2018protecting,li2019prove,namba2019robust,yang2021robust,jia2021entangled,maini2021dataset,bansal2022certified}. Specifically, during the training of the source model, the model owner utilizes a pre-selected trigger set, where the labels of the samples are known only to the owner. To verify ownership, the model owner assesses the output of the suspicious model. If the output aligns with the intended labels, the model owner can confidently assert ownership of the suspicious model.

Using a trigger set for model watermarking offers several advantages over alternative methods. Firstly, it removes the necessity for the model owner to have white-box access to the suspicious model, making it applicable in scenarios with restricted access. Secondly, it avoids the need to modify the architecture of the source model, simplifying the deployment process. However, it is crucial to note that trigger set-based watermarking methods are still susceptible to model extraction attacks \cite{lukas2022sok}. After extraction, the stolen model may fail to produce the intended labels on the trigger set, effectively removing the watermark.

Several recent efforts have sought to enhance the efficacy of trigger set-based watermarking methods \cite{li2022defending,maini2021dataset,kim2023margin}. For instance, a margin-based watermarking technique was introduced recently in \cite{kim2023margin} to optimize the margin of samples in the trigger set through projected gradient ascent. The main idea is that by emulating the decision boundary of the source model, the surrogate model can replicate predictions on the trigger set from the source model. However, this approach encounters two challenges. Firstly, the use of projected gradient ascent significantly slows the training speed of the source model. Secondly, in a more realistic attack scenario where the attacker does not leverage the source data \cite{bansal2022certified}, it is difficult for the margin-based approach to effectively 
watermark the model.

In this paper, we first introduce a novel perspective on trigger set-based watermarking methods based on feature learning. Specifically, we demonstrate the efficacy of employing multi-view data \cite{allen2020towards} as the trigger set. Multi-view data, defined as data exhibiting diverse features, is common in practice. For instance, a given horse image may possess the shape of a horse but mirror the color of a dog, exhibiting two distinct features. In model extraction attacks, we demonstrate that when the source model utilizes color as a key feature for classifying a horse as a dog, the output of the source model on this multi-view trigger set can be successfully transferred to the surrogate model. This transfer makes the removal of the watermark a challenging task. Building on this perspective, we propose a simple approach to extract multi-view data from the source data for constructing the trigger set based on logit margin. To enhance the efficacy of using the trigger set, we additionally introduce a feature regularization loss, aiming to encourage the model to learn the desired features from this specific trigger set. Aligning with recent advancements in trigger-set-based watermarking methods \cite{bansal2022certified,kim2023margin}, we conduct experiments across various widely used benchmarks. Our results demonstrate the superior performance of the proposed method compared to relevant baselines.

The contributions of the paper can be summarized as follows:
{\bf 1)} We present a novel perspective on trigger set-based watermarking methods, aiming to understand the conditions under which trigger set-based watermarking can effectively defend against model extraction attacks. To the best of our knowledge, this perspective is novel in understanding trigger set-based watermarking methods. {\bf 2)} We introduce an efficient trigger set-based watermarking approach, called MAT, by constructing a trigger set comprising multi-view data. The proposed method is simple to implement and does not require any changes to model architectures or the training process. {\bf 3)} We evaluate MAT against various attack methods and the results demonstrate that MAT enables reliable ownership identification in situations where baseline methods fail.

% \begin{itemize}
% \item We present a novel perspective on trigger set-based watermarking methods, aiming to understand the conditions under which trigger set-based watermarking can effectively defend model extraction attacks. To the best of our knowledge, this perspective is novel in understanding trigger set-based watermarking methods.

% \item We introduce an efficient trigger set-based watermarking approach, called MAT, by constructing a trigger set comprising multi-view data. The proposed method is simple to implement and does not require any changes to model architectures or the training process.

% \item We evaluate MAT against various attack methods and the results demonstrate that MAT enables reliable ownership identification in situations where baseline methods fail.
    
% \end{itemize}

\section{Related Work}
\label{sec:related}
\vspace{-0.3cm}
\noindent \textbf{Watermarking Deep Neural Networks.}
Watermarking \cite{hartung1999multimedia} was originally proposed to prove ownership of multimedia. The first model watermarking method \cite{uchida2017embedding} is proposed to protect intellectual property by adding watermarks to model parameters with a parameter regularizer during training. Based on this, more methods \cite{chen2018deepmarks,wang2019robust} have been proposed to make model watermarks more difficult to detect and remove. However, these white-box model watermarking methods usually require access to the specific parameters of the suspect model to verify whether it contains a watermark or not. This is usually difficult to do in practical scenarios.
Inspired by the backdoor attack technique \cite{gu2017badnets}, trigger set-based model watermarking methods \cite{zhang2018protecting,adi2018turning} are proposed to verify model ownership in black-box scenarios by simply querying the accuracy of a suspicious model on a specific trigger set. To make the watermark more robust, CEM \cite{lukas2019deep} uses the trigger set constructed by conferrable adversarial examples when training the model, while margin-based method \cite{kim2023margin} uses adversarial training to maximize the margins of samples from the trigger set. The use of randomized smooth \cite{chiang2020certified} during the training process has also been proven \cite{bansal2022certified} to increase the non-removable watermark of the model.

\noindent \textbf{Model Stealing.} Although model watermarking is very effective in verifying ownership, it becomes vulnerable when adversaries employ attacks to attempt to remove the watermark. In the scenario of a white-box attack, the adversary can obtain the parameter weights of the model, allowing classic defense mechanisms against backdoor attacks, such as fine-tuning and pruning \cite{liu2018fine}, to easily eliminate the model watermark. In a more realistic case, black-box attacks such as distillation \cite{hinton2015distilling} and model extraction \cite{orekondy2019knockoff} can also steal clean models without watermarks through API queries of the source model. In particular, model extraction \cite{orekondy2019knockoff} is a functionality stealing method and is regarded as the most powerful black box attack currently \cite{lukas2022sok,kim2023margin}. Some recent work \cite{bansal2022certified,kim2023margin} claims to be resistant to functionality stealing, however under some restrictions. For example, random smoothing \cite{bansal2022certified} requires limiting the size of perturbation for functionality stealing, while margin-based \cite{kim2023margin} requires the attacker to use the same training data set as the source model. Our approach effectively mitigates functionality theft in the most challenging and realistic scenarios. This includes situations where the attacker can employ an entirely different dataset, either mirroring the distribution of the source model training set or utilizing a completely unrelated dataset for model extraction.
\section{Background}
\label{sec:bg}
\vspace{-0.3cm}
\noindent \textbf{Threat model.} Similar to existing trigger set-based watermarking methods \cite{kim2023margin, bansal2022certified}, we assume that adversaries aim to emulate the functionality of the source model using surrogate data, possibly from a distribution distinct from the source dataset used to train the model. They are aware of watermarking techniques but lack knowledge of the trigger set. Meanwhile, our defense operates in a black-box scenario, where defenders seek to establish ownership of the suspected model through a trigger set. Defenders can query predicted probabilities from the suspected model but may not have access to its parameters.

\noindent \textbf{Trigger set-based approach for watermarking.}
Given a source dataset $S = { (x_i, y_i)}_{i=1}^n$ with $K$ classes, with $x_i \in \mathbb{R}^{d}$ and $y_i \in \mathbb{R}$, where both $x \in \mathcal{X}$ and $y \in \mathcal{Y}$ are sampled from a joint distribution $\mathbb{P}_{x,y}$, a specific source model $M_\theta: \mathcal{X} \rightarrow \mathbb{R}^K$ is trained by minimizing the loss function $\ell$ on the source dataset,
\begin{equation}
    \ell(M_\theta; S) = \frac{1}{|S|} \sum_{(x,y) \in S} \ell( M_\theta(x), y)
     \label{eq: source_train}
\end{equation}
Specifically, the output of the source model consists of logits represented as $M_\theta(x) = (M^1_\theta(x), M^2_\theta(x), ..., M^K_\theta(x))$, where $M^k_\theta(x)$ signifies the score assigned to class $k$ by the model.

In functionality stealing attacks, the attacker aims to train a surrogate model $\hat{M}_{\hat{\theta}}$ to imitate the functionality of the source model \cite{jia2021entangled}. In particular, the attackers leverage a surrogate data $\hat{S} = \{ (x_j, y_j)\}_{j=1}^m$ to extract the output of the source model and leverage model extraction attack for functionality stealing, 
\begin{equation}
    \min_{\hat{\theta}} \ell_{\textnormal{extraction}}(\hat{\theta}; \theta, \hat{S} ) =  \frac{1}{|\hat{S}|} \sum_{(x,y) \in \hat{S}}  D_{KL}(\hat{M}_{\hat{\theta}}(x), M_\theta(x))
    \label{eq: surrogate_train}
\end{equation}
where $D_{KL}$ is the Kullback-Leibler divergence \cite{cover1999elements} between two probability distributions. Note that the logits need to be normalized via a softmax function. Although the attacker does not have access to the source dataset or source model parameters, the functionality of the source model can still be stolen based on the output of the source model on the surrogate data \cite{orekondy2019knockoff,kim2023margin}.

\noindent \textbf{Ownership Identification with a Trigger Set.} To protect intellectual property, the owner of the source model must assess whether a particular surrogate model reproduces the functionality of the source model, a process commonly referred to as ownership identification \cite{chen2018deepmarks,quan2020watermarking,darvish2019deepsigns,yang2021robust}. One of the most commonly used ownership identification methods is based on watermarking using a trigger set \cite{adi2018turning,jia2021entangled,maini2021dataset,bansal2022certified,kim2023margin}. In particular, for training the source model, the model owner randomly samples a subset of samples $\{ (x_k, y_k)\}_{k=1}^q$ from the source dataset $S$ without replacement and replaces the true label $y_k$ with  $\hat{y}_k \neq y_k$ to construct a trigger set  $ D_{t} = \{(x_k, \hat{y}_k)\}_{k=1}^q$. Meanwhile, the trigger set is excluded from the original source dataset $S$ and the remaining clean training set is denoted as $D_{c}$ = $\{ (x_i, y_i)\}_i^m$, where \textit{q}<\textit{m}. With the additional trigger set, the model owner trains the source model by minimizing the following loss,
\begin{equation}
 \min_\theta \ell(M_\theta; D_{c}) + \ell (M_\theta; D_{t})
 \label{eq: trigger_train}
\end{equation}
Although the labels of the trigger set are not the true labels, by minimizing the above loss the source model can still achieve 100\% accuracy on the trigger set due to the high capacity of the DNNs \cite{zhang2021understanding}. As the labels of the trigger set are randomly assigned, a conventionally trained model is anticipated to attain 0\% accuracy on the trigger set when assessed using the modified labels. Should a surrogate model be trained to mimic the output of the source model, and subsequently classify the image $x_k$ in the trigger set as $\hat{y}_k$, the owner could claim that the source model has been stolen by the attacker. 

Ideally, the model owner expects the surrogate model to produce the desired label for the trigger set. However, as outlined in Eq. \ref{eq: surrogate_train}, the attack is based solely on surrogate data for functionality stealing. Consequently, there is no guarantee that the surrogate model will accurately extrapolate its predictions to the trigger set, especially considering the potential lack of correlation between the assigned labels and the images within it.

% Ideally, the model owner expects the surrogate model to produce the desired label for the trigger set. However, as outlined in Equation \ref{eq: extraction}, since the attack is only based on the surrogate data for functionality stealing. As a result, there is no guarantee that the surrogate model will accurately extrapolate its predictions to the trigger set, especially considering the lack of correlation between the labels assigned to the trigger set and the images within it. 

\section{A New Perspective for Ownership Identification}
\label{sec: perspective}
\vspace{-0.3cm}

% \begin{figure*}[t]
% \small
% \centering
% \setlength{\tabcolsep}{2pt}
%     \begin{tabular}{c c c c}
%      \includegraphics[width=0.2\linewidth]{sec/Figures/a.png}    & \includegraphics[width=0.2\linewidth]{sec/Figures/b.png} &  \includegraphics[width=0.2\linewidth]{sec/Figures/c.png}
%      & \includegraphics[width=0.2\linewidth]{sec/Figures/d.png} \\
%      a)  & b)  & c) & d) \\
%     \end{tabular}
%     \caption{a) Source training (Eq. \ref{eq: source_train}),  b) Trigger set training (Eq. \ref{eq: trigger_train}), c) Feature regularization (Eq. \ref{eq: final_loss})  d) Surrogate model (Eq. \ref{eq: surrogate_train}). The proposed MAT identifies samples close to the decision boundary as the trigger set and adjusts the features of this set to align closely with the class center of the modified label.  }
%     \label{fig:methods}
% \end{figure*}
In this paper, we introduce a novel perspective on the trigger set-based approach for watermarking based on \emph{feature learning}. This viewpoint elucidates the conditions under which the surrogate model can potentially produce the desired label on the trigger set, leading to a new way of watermarking DNNs.

Conceptually, because the attacker does not depend on the source data for extraction, transferring the predictions of the trigger set from the source model to the surrogate model can only rely on the output of the source model on the \emph{surrogate data}—a task that seems challenging. Here, we present a simple thought experiment to demonstrate the potential feasibility of such a transfer. Suppose we have an image of a horse that shares certain attributes, such as color, with those of a dog. If we select this image as the trigger set and assign the label "dog" to it, we create a scenario where the attributes of a horse are mislabeled as those of a dog. During the training of the source model using Eq. \ref{eq: source_train}, the model is likely to rely on color cues in its classification process to identify the image as a dog. In this specific scenario, color serves as the exclusive distinguishing feature for classifying the image into the dog category. Now, in the context of a model extraction attack, let's assume there exists a dog image within the surrogate dataset that shares the same color as the horse image in the trigger set. The source model, when presented with this dog image, would predict it as a dog, and this prediction would transfer to the surrogate model based on Eq. \ref{eq: surrogate_train}. As a result, the surrogate model, having adopted this prediction, would subsequently categorize the horse image within the trigger set as a dog based on the color cue. To make the above idea concrete, we will leverage the following definition of multi-view data,

\noindent \textbf{Multi-view data.} The multi-view hypothesis \cite{allen2020towards} states that a given sample may exhibit multiple distinct features which can be used for classification. In the example above, the horse image exhibits both a horse's distinctive shape characteristics and a dog's color attributes. Suppose we have two classes: dog and horse, for simplicity. We assume each class $c$ has one feature $v_{c} \in \mathbb{R}^{p \times 1}$ which represents the mean feature of this class. We further assume the feature vectors of different classes are orthogonal, \ie, $v_{0}^Tv_{1} =0$. For the sample $x^c_i$ belonging to class $c$, if the sample exhibits features from both classes, then we can represent the feature of this sample as $f_i^c = w^0_i v_{0} + w^1_i v_{1}$, where $w^0_i, w^1_i \in \mathbb{R}$ are some constant weights. For example, a given horse image may resemble a dog such that the feature of the horse can be written as $f_{horse} \approx 0.2v_{0} + 0.8v_{1}$. As shown in \cite{allen2020towards}, multi-view data commonly exists in real-world datasets, as natural images often possess diverse features that can be exploited for classification.

\noindent \textbf{Watermarking using multi-view data.} Here we show that multi-view data can be naturally used for constructing the trigger set. Assume that we assign a dog class to the horse image. Consider a simple binary classifier with weights $y =Wx + b$, where $W \in \mathbb{R}^{2 \times p}$ and $b \in \mathbb{R}^2$. We use $W_i$ to denote the $i$-th row of $W$. Since the trigger set is trained alongside the clean data and is relatively small in comparison, it is reasonable to assume that the classifier can perfectly classify the two classes and the weights are perfectly aligned with the corresponding class, that is, $ \cos (W_{i}, v_{i}) = 1$. Given the sample $x^1_i$ with feature $f_i^1 = w^0_i v_{0} +  w^1_i v_{1}$, the logit can be computed as,
\begin{equation}
\begin{split}
 z & = W f_i^1 + b = w^0_i*W  v_0 +  w^1_i*W v_1 + b \\
 & = \begin{bmatrix}
w^0_iW_0v_0 + w^1_iW_0v_1 + b_0\\
w^0_iW_1v_0 + w^1_iW_1v_1 + b_1
\end{bmatrix}    
\end{split}
\end{equation}
The logit will be converted into probabilities using a softmax function,
\begin{equation}
\text{softmax}(z)_k = \frac{e^{z_k}}{\sum_{j=1}^{2} e^{z_j}}, \quad k = 0, 1
\end{equation}

If we assign a class 0 to this image, the binary cross-entropy loss can be simplified to $\ell = -\log(\text{softmax}(z)_0)$. Based on our assumptions, $\text{softmax}(z)_0 = \frac{e^{w^0_iW_0v_0 + b_1}}{\sum_{j=1}^{2} e^{z_j}}$ as $W_0v_1 = 0$. Thus the source model will leverage the feature $v_0$ for classifying the images. When a sample with feature $v_0$ is provided in a model extraction attack, the source model predicts class 0. This prediction is then leveraged to train a surrogate model, resulting in the surrogate model predicting the same label for the sample within the trigger set.
% \vspace{-0.3cm}
\section{MAT}
\label{sec: mat}
\vspace{-0.3cm}
Based on the above analysis, we introduce MAT, a novel watermarking method using trigger sets to defend against model extraction attacks. MAT incorporates a novel trigger sample selection method and a new approach to source model training. In the Appendix, we illustrate our entire training pipeline.

\begin{figure}[t]

\centering
\setlength{\tabcolsep}{8pt}
\begin{tabular}{ccccc}
{ \scriptsize Bird (Airplane)} & { \scriptsize Frog (Deer)} & { \scriptsize Dog (Cat)} & { \scriptsize Dog (Cat)} & { \scriptsize Dog (Horse)} \\
  \includegraphics[]{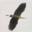}   &  \includegraphics[]{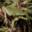} &  \includegraphics[]{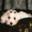}  &  \includegraphics[]{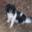}  &  \includegraphics[]{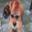}  \\
\end{tabular}
\caption{Some sample images in the trigger set selected by MAT on CIFAR-10. The images exhibit features from different classes as expected. The true classes, along with the classes having the second-highest scores, are displayed in parentheses.}
\label{fig: images}
\end{figure}

\noindent \textbf{Margin-based Trigger Set Selection.} While our analysis in Section \ref{sec: perspective} demonstrates the effectiveness of transferring the source model's predictions on the trigger set to the surrogate model through the utilization of data with features from different classes, it remains unclear how to identify samples exhibiting multi-view features within a given source dataset. To tackle this challenge, we introduce a simple and effective method for extracting multi-view data from the source dataset and subsequently employ them to construct the trigger set.

Given a source model $M_\theta$ and the logits $M_\theta(x)$ for a sample $x$ with label $y$, if the sample exhibits features from different classes, then the model will assign high scores to the respective classes. This intuition can be captured by the logit margin loss which is defined as, $LM = \max_{j \neq y} M^j_\theta(x) - M^y_\theta(x)$. If the model makes the correct prediction, a large logit margin loss indicates that the features of sample $x$ comprise characteristics of both the true classes and the class with the second-highest score. Thus, we can select a trigger set consisting of samples with large logit margin loss. In practice, we first train the model on the source dataset $S$ using Eq. \ref{eq: source_train} and then select the top $q$ samples with the largest logit margin loss as the trigger set. The label of each selected sample is \textit{modified} to the class $\hat{y} = \argmax_{j \neq y} M^j_\theta(x) - M^y_\theta(x)$. Fig. \ref{fig: images} shows the selected images using the proposed margin-based trigger set selection on CIFAR-10. It can be observed that the images are difficult to classify and exhibit features from different classes.

\begin{figure*}[t]
\small
\centering
\setlength{\tabcolsep}{2pt}
    \begin{tabular}{c c c c}
     \includegraphics[width=0.2\linewidth]{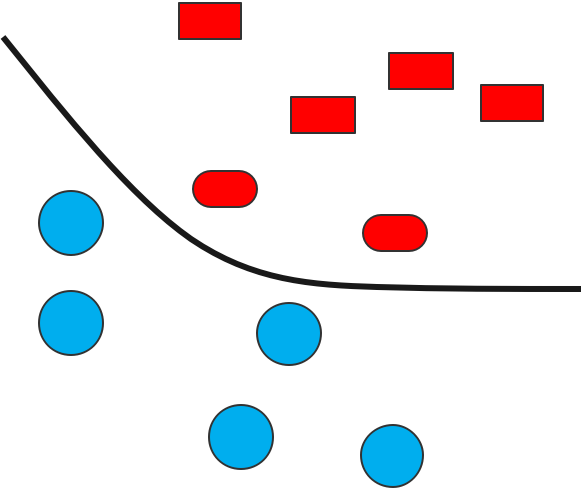}    & \includegraphics[width=0.2\linewidth]{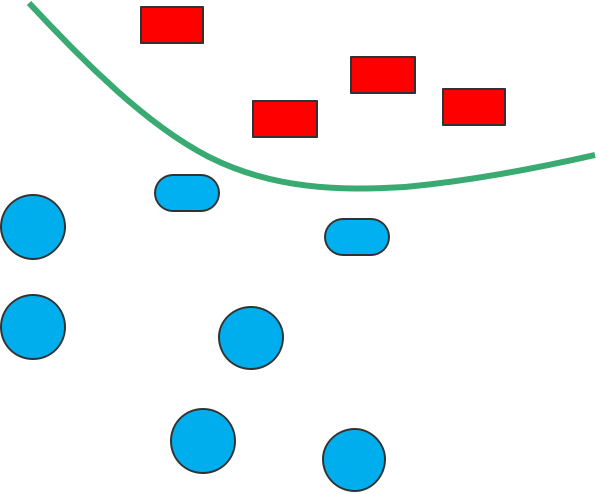} &  \includegraphics[width=0.2\linewidth]{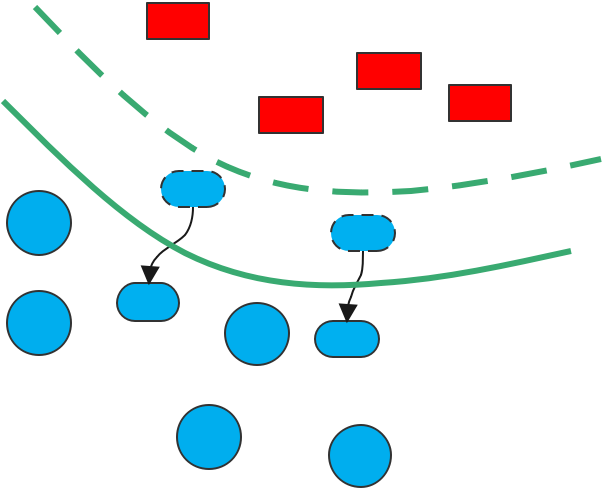}
     & \includegraphics[width=0.2\linewidth]{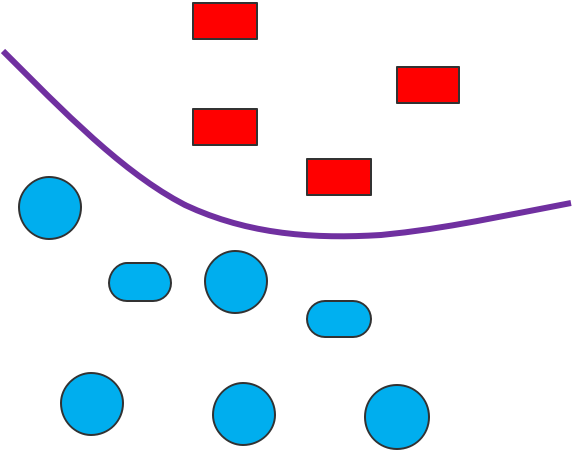} \\
     a)  & b)  & c) & d) \\
    \end{tabular}
    \caption{a) Source training (Eq. \ref{eq: source_train}),  b) Trigger set training (Eq. \ref{eq: trigger_train}), c) Feature regularization (Eq. \ref{eq: final_loss})  d) Surrogate model (Eq. \ref{eq: surrogate_train}). The proposed MAT identifies samples close to the decision boundary as the trigger set and adjusts the features of this set to align closely with the class center of the modified label.  }
    \label{fig:methods}
\end{figure*}

In Fig. \ref{fig:methods}, we show a simple example with two classes: circle and rectangle. With the logit margin loss, we select samples close to the decision boundary that exhibit features of both the circle and rectangle classes. Fig. \ref{fig:methods} a) shows the decision boundary of the source model
on the source data. Fig. \ref{fig:methods} b) shows the decision boundary of the source model after training with the clean data and the chosen trigger set. It is noteworthy that this selection not only enhances watermarking performance but also imposes minimal effect on clean accuracy, given that all other samples from the rectangle class are considerably far from the chosen trigger set. This intuition is validated by the experimental results.

\noindent \textbf{Feature Regularization.} While the chosen trigger set possesses multi-view features, it is still important to encourage the source model to learn the desired features effectively. In Fig. \ref{fig:methods} b), ideally, the source should employ the circle feature to classify samples in the trigger set. For instance, in a model extraction attack involving a circular image, the source model should be capable of transferring predictions from the trigger set to the surrogate model. Thus, we propose to enhance the model's feature learning by incorporating a feature regularization loss. In particular, for a given sample $(x_k, \hat{y}_k)$ in the trigger set, we aim to push the feature of the sample to be close to the mean feature of class $\hat{y}_k$. This can be achieved by minimizing the following loss,
\begin{equation}
\small
 \min_\theta \ell(\theta; D_{c}) + \ell (\theta; D_{t}) + \alpha \frac{1}{|D_{t}|} \sum_{(x_k,  \hat{y}_k) \in D_{t}} \| f(x_k) - f_{\hat{y}_k} \|_2
 \label{eq: final_loss}
\end{equation}
where $\alpha$ is a balance parameter. As the mean feature of the class $\hat{y}_k$ is not directly available, we employ an approximation by computing them through the average features of samples belonging to the class $\hat{y}_k$. Specifically, we calculate the average features for each class during the previous epoch of model training and utilize these averages to regularize the feature learning of the trigger set in the current epoch. Fig. \ref{fig:methods} c) demonstrates that additional feature regularization pushes trigger set features towards the circle class center, shifting the source model's decision boundary accordingly. In Fig. \ref{fig:methods} d), the surrogate model's decision boundary after a model extraction attack aligns with the source model's predictions on the trigger set, as it aims to replicate them on the surrogate dataset. We include the discussions and results of an alternative regularization method in the Appendix.

\emph{Crucially, MAT does not require the surrogate data to exhibit the features of the source dataset to successfully identify ownership.} This is because the output of the source model for any given input can be used to approximate the decision boundary of the source model. By shaping the decision boundary of the source model using the chosen trigger set, the prediction of the source model on the trigger set can be transferred to the surrogate model. We experimentally validate this in Sec. \ref{sec: abalation}.
% \vspace{-0.3cm}

\noindent \textbf{Ownership Verification with Hypothesis Test.} Our pipeline for ownership verification is similar to existing watermarking-based methods \cite{li2022untargeted,li2023black, guo2024domain, kim2023margin}. The defender can assert ownership of suspicious models by demonstrating statistically significant deviations in their behavior compared to a benign model. Specifically, we train a model $M^c_{\theta}$ directly on the clean dataset $D_c$, which serves as our benign model. We then evaluate the predictions of both the benign model $M^c_{\theta}$ and the suspicious model $\hat{M}_{\hat{\theta}}$ on the trigger set $D_t$, denoting their predictions as $P$ and $\hat{P}$, respectively. These predictions can be represented as a binomial distribution, which indicates whether watermarked data is classified as the target class\cite{jia2021entangled}. Validation is conducted through a two-sample t-test, comparing the models' predictions on the trigger set. If the resulting p-value is less than the significance level, defenders can claim ownership of the model. 

\iffalse
\vspace{-0.2cm}
 \begin{figure}[htb!]
    \centering
\vspace{-0.4cm}
\includegraphics[width=0.38\textwidth]{sec/Figures/pipeline.png}
\vspace{-0.2cm}
    \caption{Pipeline of our method.}
    \label{fig:trigger}
\end{figure}
\vspace{-0.2cm}
\fi

%given a trigger set $D_t$ and a model $M_{source}$ that utilizes it to add watermarking, we first train the $model_{surrogate}$ on another clean dataset $D_c$ by extracting attacks,  and then train the $model_{clean}$ directly on $D_c$ as a comparison. We compute the p-value by comparing the binomial distribution of whether these two models predict success on trigger set $D_t$.
\section{Experiments}
\vspace{-0.3cm}
\subsection{Experimental Settings}
\vspace{-0.1cm}
\noindent \textbf{Datasets.} Building upon prior studies \cite{bansal2022certified, kim2023margin}, we utilize the CIFAR-10 \cite{krizhevsky2009learning} and CIFAR-100 \cite{krizhevsky2009learning} datasets. Additionally, we extend our evaluation to include ImageNet \cite{deng2009imagenet}, a large-scale image dataset that challenges existing watermarking techniques. Following \cite{bansal2022certified}, we use half of the training dataset as the source data, with the remaining half serving as the surrogate dataset.

\noindent \textbf{Baselines.} We compare MAT with the following baseline approaches:
\begin{itemize}[noitemsep,topsep=0pt,parsep=0pt,partopsep=0pt]
    \item  \textbf{Base} \cite{zhang2018protecting}. Randomly selects a trigger set and trains the model using Eq. \ref{eq: trigger_train}.

    \item  \textbf{Randomized Smoothing (RS)} \cite{bansal2022certified}. It employs randomized smoothing to achieve certified robustness when the modifications to the source model parameters $\theta$ are limited in size.

    \item \textbf{Margin-based method} \cite{kim2023margin}. It maximizes the margin of the samples in the trigger set using the projected gradient descent. 
    
    \item  \textbf{Datasets Inference (DI)} \cite{maini2021dataset}. It verifies ownership by determining whether the suspected model contains private knowledge of the source model's training dataset. 
    \item  \textbf{Embedded External Features (EEF)} \cite{li2022defending}. It transfers the data style of the trigger set by embedding external features.

\end{itemize}

\noindent \textbf{Metrics.} We evaluate the accuracy of the surrogate model $\hat{M}_{\hat{\theta}}$ on the trigger set $D_t$, and report the p-value using a two-sample t-test, where a small p-value indicates distinguishability between the stolen and benign models. Additionally, we report the clean accuracy of both the source and surrogate models.

\noindent \textbf{Attack methods.} We consider several commonly used attack methods:
\begin{itemize}[noitemsep,topsep=0pt,parsep=0pt,partopsep=0pt]
    \item \textbf{Soft-label model extraction attack}. The attacker leverages the predictions of the source model on the surrogate model as in Eq. \ref{eq: surrogate_train}.
    \item \textbf{Hard-label model extraction}. Instead of depending on the soft labels $M_\theta(x) = (M^1_\theta(x), M^2_\theta(x), ..., M^K_\theta(x))$, the attacker exploits the hard label $y_{hard} = \argmax_k{M^c_\theta(x)}_{k=1}^K$. The hard label will be converted into a one-hot representation, which then replaces the soft label in Eq. \ref{eq: surrogate_train} for a model extraction attack.
    \item \textbf{Fine-tuning} \cite{liu2018fine}. Given the access to the source model, the attacker fine-tunes the source model on a clean data set under the same data distribution.   
    \item \textbf{Fine-pruning} \cite{liu2018fine}. Given access to the source model, the attacker prunes the neurons in the last convolutional layer based on their activation levels using a small batch of clean data and then proceeds with fine-tuning.
\end{itemize}

\noindent \textbf{Implementation details.} 
We utilize ResNet-18 \cite{he2016deep} as the source and surrogate models for CIFAR-10 and CIFAR-100, and ViT-base-patch16-384 \cite{dosovitskiy2020image} for ImageNet. Additional results with alternative architectures are provided in the Appendix. We assume that all surrogate models are randomly initialized, as attackers do not have access to the parameters of the source model. In all the experiments, the size of the trigger set is 100. We first train a clean model on the source data for 200 epochs for extracting multi-view data for constructing the trigger set. Then, a source model is trained on the clean data and the trigger set data as in Equation \ref{eq: final_loss}. We follow the same training strategies as in  \cite{kim2023margin}. The source model is trained for 200 epochs and optimized using SGD \cite{kiefer1952stochastic} with an initial learning rate of 0.1. A linear decay is used to decay the learning rate every 50 epochs. For the details of various attack methods, please see the Appendix.
\vspace{-0.2cm}
\subsection{Results on Functionality Stealing}
\vspace{-0.2cm}
Tables \ref{tab:cifar10_res}, \ref{tab:cifar100_res} and \ref{tab:imagenet_res} present the source accuracy (Source Acc.), surrogate accuracy (Surro. Acc.), trigger set accuracy (Trig. Acc) and p-value for all the methods on CIFAR-10, CIFAR-100, and ImageNet, respectively. \emph{For the statistical testing based methods - DI \cite{maini2021dataset} and EEF \cite{li2022defending}, we report only the p-values}. It is evident that following a model extraction attack, all the baselines exhibit a substantial number of errors on the trigger set, posing a significant challenge to ownership verification. Specifically, in the case of a hard-label model extraction attack, none of the baselines can transfer the prediction of the source model on the trigger set to the surrogate model, rendering ownership verification impossible.

\begin{table*}[!t]
\caption{Our proposed MAT achieves the best watermarking performance on CIFAR-10 compared with the baselines. Even after the hard-label model extraction attack, MAT still achieves 56\% accuracy on the trigger set.}
    \centering
\scriptsize
\resizebox{\textwidth}{!}{%
    \begin{tabular}{c|c|c|c|c|c|c|c}
    \toprule

         & \multirow{2}{*}{Source Acc. (\%)}   & \multicolumn{3}{c}{Soft-Label}  & \multicolumn{3}{c}{Hard-Label} \\

         &   & Surro. Acc. (\%)& Trig. Acc. (\%) & p-Value & Surro. Acc. (\%) & Trig. Acc. (\%) & p-Value\\
         \midrule

        Base \cite{zhang2018protecting} &  90.40 & 89.07&  0 &$10^{-1}$ & 85.22  &4& $10^{-1}$\\
            \midrule
        RS \cite{bansal2022certified}  &  91.10 &89.74 & 2&$10^{-1}$ & 85.75&2&$10^{-1}$\\
             \midrule
       Margin-based \cite{kim2023margin} & 83.80&  86.51 & 46 &$10^{-14}$& 84.45 & 10&$10^{-3}$\\
            \midrule
       DI \cite{maini2021dataset} &-&-&-&$10^{-6}$&-&-&$10^{-3}$\\
    \midrule
       EEF \cite{li2022defending} &-&-&-&$10^{-7}$&-&-&$10^{-4}$\\
    \midrule
        \midrule
    MAT (no reg.)  & 91.70 & 89.44 & 52 &$10^{-4}$&86.05 & 37&$10^{-1}$\\
    \midrule

   MAT & 87.90 & 88.50 & \textbf{74} &$10^{-11}$& 85.40& \textbf{56}&$10^{-4}$  \\
    \bottomrule
    \end{tabular}}
    \label{tab:cifar10_res}
\end{table*}

\begin{table*}[t]
\caption{Our proposed MAT achieves the best watermarking performance on CIFAR-100 compared with the baselines after the soft-label model extraction attack. With the hard-label model extraction attack, the attack itself becomes more challenging, yet MAT still achieves higher trigger accuracy compared to other methods.}
    \centering
\scriptsize
\resizebox{\textwidth}{!}{%
    \begin{tabular}{c|c|c|c|c|c|c|c}
    \toprule

         & \multirow{2}{*}{Source Acc. (\%) }   & \multicolumn{3}{c}{Soft-Label}  & \multicolumn{3}{c}{Hard-Label} \\
         &   & Surro. Acc. (\%)& Trig. Acc. (\%) & p-Value & Surro. Acc. (\%) & Trig. Acc. (\%) & p-Value\\
         \midrule
         Base \cite{zhang2018protecting} & 64.30&63.99&2&$10^{-1}$&15.01&1&$10^{-1}$\\
         \midrule
         RS \cite{bansal2022certified}  &  66.80 &64.85 & 0 &$10^{-1}$& 15.61&0&$10^{-1}$ \\
             \midrule
       Margin-based \cite{kim2023margin} & 60.95  & 61.33 & 11&$10^{-3}$& 15.74 & 1&$10^{-1}$\\
    \midrule
    DI \cite{maini2021dataset} &-&-&-&$10^{-2}$&-&-&$10^{-1}$\\
    \midrule
     MAT (no reg.) &66.02 & 65.65  & 62&$10^{-11}$& 15.44& \textbf{9}&$10^{-1}$\\
    \midrule

   MAT  &61.20  & 59.73 & \textbf{77}&$10^{-20}$ & 15.19& \textbf{9}&$10^{-1}$\\
    \bottomrule
    \end{tabular}}
    \label{tab:cifar100_res}
\end{table*}

In contrast, the proposed MAT effectively retains the majority of predictions on the trigger set. Specifically, when compared to the margin-based method, the trigger set accuracy experiences a substantial improvement, rising from 46\% to 74\%. Even when subjected to a hard-label model extraction attack, MAT still maintains a trigger set accuracy of 56\% on CIFAR-10. On CIFAR-100, all baseline methods fail to maintain trigger set accuracy following a soft-label model extraction attack, whereas our proposed MAT demonstrates an accuracy of 77\%. Similarly, MAT achieves much higher trigger set accuracy and a lower p-value on ImageNet, demonstrating its scalability to large image datasets.

Consistent with the findings in \cite{bansal2022certified}, we observe that all methods struggle to preserve trigger set accuracy following a hard-label model extraction attack. This phenomenon may be attributed to the inherent challenges of the CIFAR100 dataset and ImageNet, coupled with the low accuracy of the source model on the surrogate dataset. We also present results for MAT (no reg.), which does not use feature regularization during training. MAT (no reg.) achieves higher clean accuracy but lower trigger set accuracy compared to MAT. Thus, MAT (no reg.) can be employed when prioritizing high clean accuracy over ownership verification.

\begin{table*}[t]
\caption{Our proposed MAT achieves the best watermarking performance on ImageNet compared with the baselines. }
    \centering
\scriptsize
\resizebox{\textwidth}{!}{%
    \begin{tabular}{c|c|c|c|c|c|c|c}
    \toprule

         & \multirow{2}{*}{Source Acc. (\%)}   & \multicolumn{3}{c}{Soft-Label}  & \multicolumn{3}{c}{Hard-Label} \\

         &   & Surro. Acc. (\%)& Trig. Acc. (\%) & p-Value & Surro. Acc. (\%) & Trig. Acc. (\%) & p-Value\\
         \midrule

        Base \cite{zhang2018protecting} &  75.30 & 74.44 &  1 &$10^{-1}$ & 5.84  &4& $10^{-1}$\\
            \midrule
       Margin-based \cite{kim2023margin} & 70.06 & 71.92 & 9 &$10^{-2}$& 5.68 & 1 &$10^{-1}$\\
    \midrule
    \midrule
   MAT (no reg.) & 74.94 &  72.76 & \textbf{82} &$10^{-7}$& 5.36 & \textbf{10}&$10^{-1}$  \\
        \midrule
   MAT & 76.25 &  74.16 & \textbf{82} &$10^{-7}$& 5.84 & \textbf{6}&$10^{-1}$  \\
    \bottomrule
    \end{tabular}}
    \label{tab:imagenet_res}
\end{table*}
\vspace{-0.3cm}
\subsection{Results with White-box Attacks}
\begin{figure}[t]
\centering
\setlength{\tabcolsep}{1pt}
\begin{tabular}{cc}
  \includegraphics[width=0.4\linewidth]{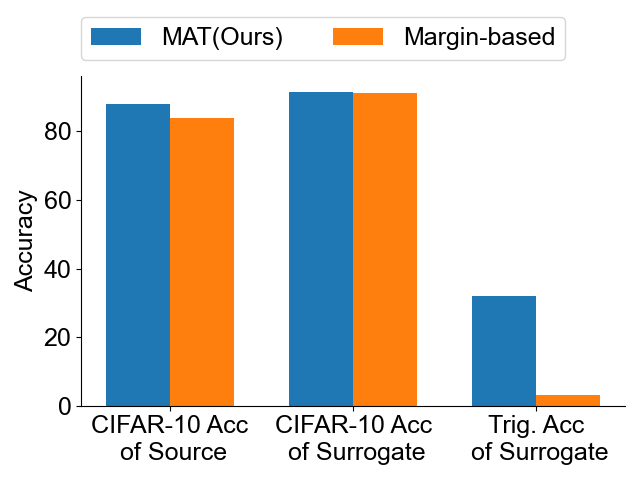}   &  \includegraphics[width=0.4\linewidth]{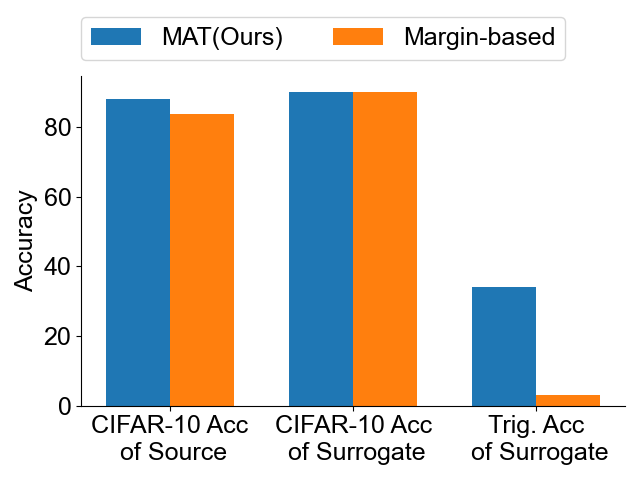}   \\
 a) Fine-tuning & b) Fine-pruning
\end{tabular}
    \caption{MAT significantly outperforms the margin-based approach in watermarking effectiveness when facing white-box attacks.}
    \label{fig:whitebox}
\end{figure}

\vspace{-0.3cm}
A model extraction attack is a black-box attack that assumes that the attacker cannot have access to the parameters of the source model. In some cases, the attacker may even know the parameters of the source model in which case the attacker can perform a white-box attack. In this section, we examine two white-box attacks, namely fine-tuning and fine-pruning. As illustrated in Fig. \ref{fig:whitebox}, our proposed MAT demonstrates the ability to sustain a reasonably high trigger set accuracy, even when subjected to white-box attacks. This represents a significant strength of MAT in contrast to margin-based approaches, which lack the capability to effectively defend against such attacks.

\vspace{-0.2cm}
\subsection{Ablation Studies}
\vspace{-0.2cm}
\label{sec: abalation}
\noindent \textbf{Heterogeneous Surrogate Dataset \& Model.} MAT does not need to assume that the surrogate data possess the features of the source dataset. In this section, we address a more realistic scenario where the attacker lacks knowledge of the distribution of the source data or the architecture of the source model. Instead of the CIFAR-10 dataset, we employ the SVHN dataset \cite{netzer2011reading} as the surrogate dataset, maintaining ResNet-18 as the architecture for the surrogate model. Additionally, we explore the use of the CIFAR-10 dataset as the surrogate dataset while employing VGG11 \cite{simonyan2014very} as the architecture for the surrogate model. Table \ref{tab:hetero} reports that in both cases, the proposed MAT consistently outperforms the margin-based approach in terms of trigger set accuracy. Visualization results using t-SNE \cite{van2008visualizing} can be found in the Appendix.

\begin{table}[t]
\caption{MAT still outperforms the margin-based approach with heterogeneous surrogate data and architecture. }
\scriptsize
    \centering
    \begin{tabular}{c|c|c}
      \toprule
      Method   & Surrogate Clean Acc. (\%)  & Trigger Set Acc. (\%) \\
             \midrule
        \multicolumn{3}{c}{ \textbf{Surrogate Dataset with SVHN}} \\
         \midrule
       Margin-based \cite{kim2023margin}  & 82.53 & 39  \\
             \midrule
    MAT  & 82.43 & \textbf{67} \\
         \midrule
                 \midrule
 \multicolumn{3}{c}{ \textbf{Surrogate Model with VGG11} } \\
         \midrule
      Margin-based \cite{kim2023margin}  & 85.69 & 43 \\
             \midrule
    MAT  & 85.77 & \textbf{62} \\
    \bottomrule      
    \end{tabular}
    \label{tab:hetero}
\end{table}

\noindent \textbf{Different trigger set selection strategies.} In this section, we further demonstrate the effectiveness of the proposed trigger set selection strategy in comparison to alternative approaches. Specifically, we examine the following strategies:
\begin{itemize}[noitemsep,topsep=0pt,parsep=0pt,partopsep=0pt]
\item Random selection (Random): This strategy randomly selects a subset of samples from the source dataset to form the trigger set.
\item Highest confidence (Highest conf.): This strategy selects samples with the smallest logit margin loss over those with the largest.
\end{itemize}
For both of these strategies, we maintain consistency with the labeling strategy employed by MAT. The results presented in Table \ref{tab:trigger_set_selection_cifar10} and \ref{tab:trigger_set_selection_cifar100} show the performance of these strategies alongside MAT. With the adoption of the proposed trigger set selection strategy, MAT significantly outperforms the other two approaches, emphasizing the effectiveness of choosing multi-view data as the trigger set.

\begin{minipage}[c]{0.47\textwidth}
\captionof{table}{Results of trigger set selection strategies on CIFAR-10.}
\scriptsize
\centering
    \begin{tabular}{c|c|c|c}
             \toprule
         & \multicolumn{2}{c|}{Clean Acc.}    & \multirow{2}{*}{Trig. Acc. (\%)}\\
         & Source & Surrogate & \\
       \midrule
       Random  & 87.73 &  87.74& 12\\
       \midrule
       Highest conf. & 82.68  &  83.80 &  1 \\
       \midrule
       MAT & 87.90  & 88.50  & \textbf{74} \\
       \bottomrule
    \end{tabular}   
\label{tab:trigger_set_selection_cifar10} 
\end{minipage}
\hfill
\begin{minipage}[c]{0.47\textwidth}
\captionof{table}{Results of trigger set selection strategies on CIFAR-100}
\scriptsize
\centering
    \begin{tabular}{c|c|c|c}
             \toprule
         & \multicolumn{2}{c|}{Clean Acc.}    & \multirow{2}{*}{Trig. Acc. (\%)}\\
         & Source & Surrogate & \\
       \midrule
       Random  & 59.14 & 58.97 & 56\\
       \midrule
       Highest conf. & 58.75 & 59.23  & 9\\
       \midrule
       MAT &61.20   & 59.73  & \textbf{77} \\
       \bottomrule
    \end{tabular}
    \label{tab:trigger_set_selection_cifar100}
\end{minipage}

\iffalse

\begin{table}[h]
    \centering
    \scriptsize
    \begin{tabular}{c|c|c|c}
             \toprule
         & \multicolumn{2}{c|}{Clean Acc.}    & \multirow{2}{*}{Trigger Set Acc. (\%)}\\
         & Source & Surrogate & \\
       \midrule
       Random selection  & 87.73 &  87.74& 12\\
       \midrule
       Highest conf. & 82.68  &  83.80 &  1 \\
       \midrule
       MAT & 87.90  & 88.50  & \textbf{74} \\
       \bottomrule
    \end{tabular}
    \caption{Results of trigger set selection strategies on CIFAR-10.}
    \label{tab:trigger_set_selection_cifar10}
\end{table}

\begin{table}[h]
    \centering
    \scriptsize
    \begin{tabular}{c|c|c|c}
             \toprule
         & \multicolumn{2}{c|}{Clean Acc.}    & \multirow{2}{*}{Trigger Set Acc. (\%)}\\
         & Source & Surrogate & \\
       \midrule
       Random selection & 59.14 & 58.97 & 56\\
       \midrule
       Highest conf. & 58.75 & 59.23  & 9\\
       \midrule
       MAT &61.20   & 59.73  & \textbf{77} \\
       \bottomrule
    \end{tabular}
    \caption{Results of trigger set selection strategies on CIFAR100.}
    \label{tab:trigger_set_selection_cifar100}
\end{table}

\fi

\vspace{0.3cm}

\noindent \textbf{Different trigger set labeling strategies.} In addition to choosing the trigger set, how to modify the label of the trigger set is also critical for transferring the prediction of the source model on the trigger set to the surrogate model. To further show the effectiveness of our labeling strategy, we examine the following two alternatives:
\begin{itemize}[noitemsep,topsep=0pt,parsep=0pt,partopsep=0pt]
    \item Random labeling (Random): Randomly assign one of the labels, excluding the true label, to the selected sample.
      \item Minimize confidence (Min. conf.): Identify the label that minimizes the logit margin loss. The label can be calculated as $\Tilde{y} = \argmin_{j \neq y} \left( M^j_\theta(x) - M^y_\theta(x) \right)$. 
\end{itemize}
For both of these strategies, we maintain consistency with the trigger set selection strategy employed by MAT. The results presented in Table \ref{tab:trigger_set_labelling_cifar10} and \ref{tab:trigger_set_labelling_cifar100} showcase the performance of these strategies alongside MAT. With the adoption of the proposed trigger set labelling strategy, MAT significantly outperforms the other two approaches. In particular, MAT assigns the label to the sample in the trigger that the model finds most confusing, making it easier for the model to learn the distinctive features of the assigned classes. Consequently, it becomes feasible to transfer the model's predictions on the trigger set to the surrogate model.

\begin{minipage}[c]{0.47\textwidth}
\captionof{table}{Results of trigger set labelling strategies on CIFAR-10.}
\scriptsize
\centering
    \begin{tabular}{c|c|c|c}
             \toprule
         & \multicolumn{2}{c|}{Clean Acc. (\%)}    & \multirow{2}{*}{Trig. Acc. (\%)}\\
         & Source & Surrogate & \\
       \midrule
       Random  &85.66 & 85.82 & 32 \\
       \midrule
       Min conf.  & 86.92  & 86.91  & 2 \\
       \midrule
       MAT & 87.90  & 88.50  &  \textbf{74}\\
       \bottomrule
    \end{tabular} 
\label{tab:trigger_set_labelling_cifar10} 
\end{minipage}
\hfill
\begin{minipage}[c]{0.47\textwidth}
\captionof{table}{Results of trigger set labelling strategies on CIFAR-100.}
\scriptsize
\centering
    \begin{tabular}{c|c|c|c}
             \toprule
         & \multicolumn{2}{c|}{Clean Acc. (\%)}    & \multirow{2}{*}{Trig. Acc. (\%)}\\
         & Source & Surrogate & \\
       \midrule
       Random  & 57.43 & 58.45 & 15\\
           \midrule
       Min conf.  & 55.94 & 56.61 & 9 \\
       \midrule
       MAT & 61.20  & 59.73  &  \textbf{77}\\
       \bottomrule
    \end{tabular}
    \label{tab:trigger_set_labelling_cifar100}
\end{minipage}

\iffalse

\begin{table}[h]
    \centering
    \scriptsize
    \begin{tabular}{c|c|c|c}
             \toprule
         & \multicolumn{2}{c|}{Clean Acc.}    & \multirow{2}{*}{Trigger Set Acc. (\%)}\\
         & Source & Surrogate & \\
       \midrule
       Random labeling  &0.8566 & 0.8582 & 32 \\
       \midrule
       Min conf.  & 0.8692  & 0.8691  & 2 \\
       \midrule
       MAT & 87.90  & 88.50  &  \textbf{74}\\
       \bottomrule
    \end{tabular}
    \caption{Results of trigger set labelling strategies on CIFAR10.}
    \label{tab:trigger_set_labelling_cifar10}
\end{table}

\begin{table}[h]
    \centering
    \scriptsize
    \begin{tabular}{c|c|c|c}
             \toprule
         & \multicolumn{2}{c|}{Clean Acc.(\%)}    & \multirow{2}{*}{Trigger Set Acc. (\%)}\\
         & Source & Surrogate & \\
       \midrule
       Random labeling  & 57.43 & 58.45 & 15\\
           \midrule
       Min conf.  & 55.94 & 56.61 & 9 \\
       \midrule
       MAT & 61.20  & 59.73  &  \textbf{77}\\
       \bottomrule
    \end{tabular}
    \caption{Results of trigger set labeling strategies on CIFAR-100.}
    \label{tab:trigger_set_labelling_cifar100}
\end{table}
\fi

\vspace{0.4cm}

\begin{figure}[h]
\vspace*{-0.5cm}
\centering
\small
\begin{minipage}{.48\textwidth}
\begin{tabular}{cc}
  \includegraphics[width=0.5\linewidth]{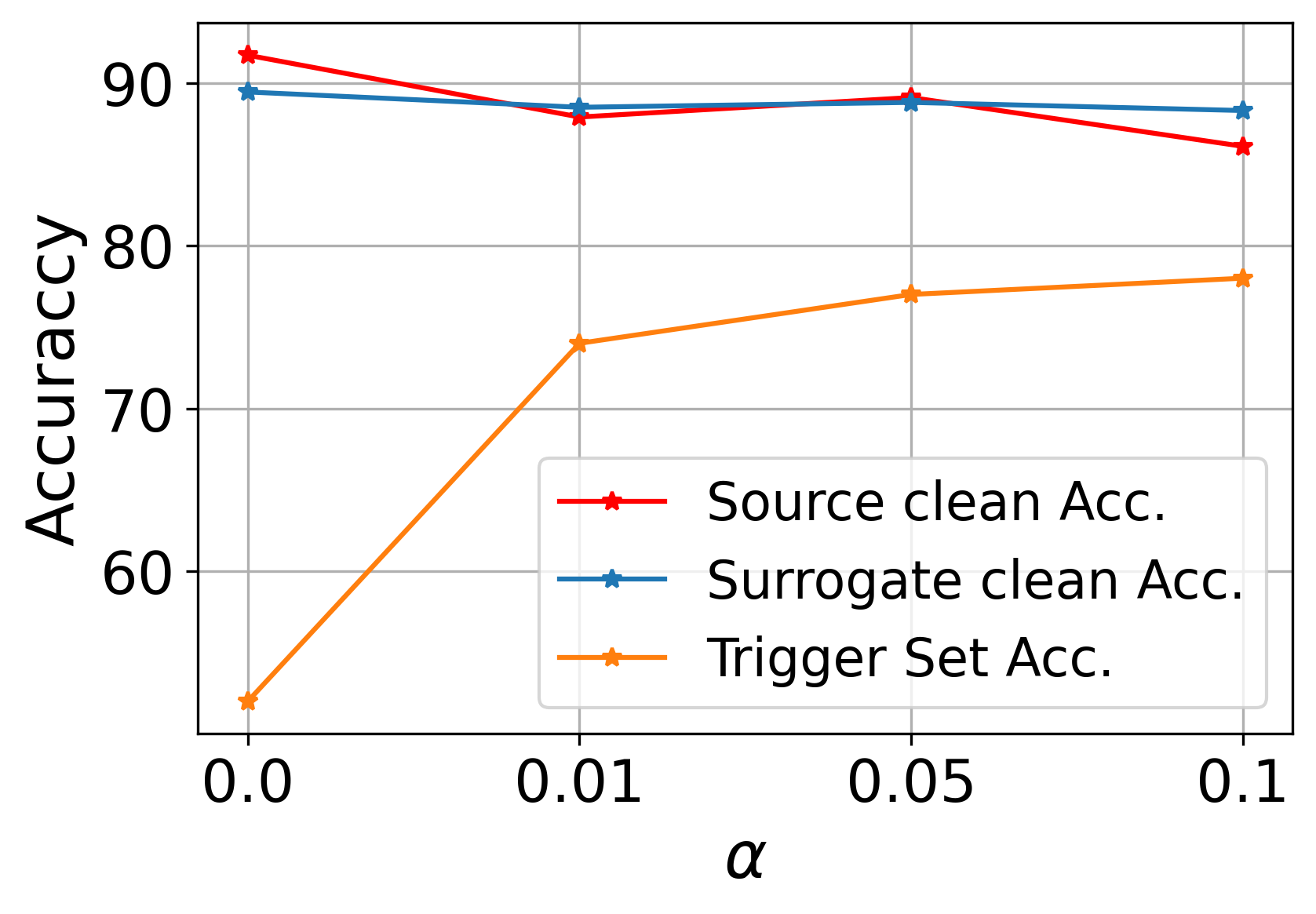}   &  \includegraphics[width=0.5\linewidth]{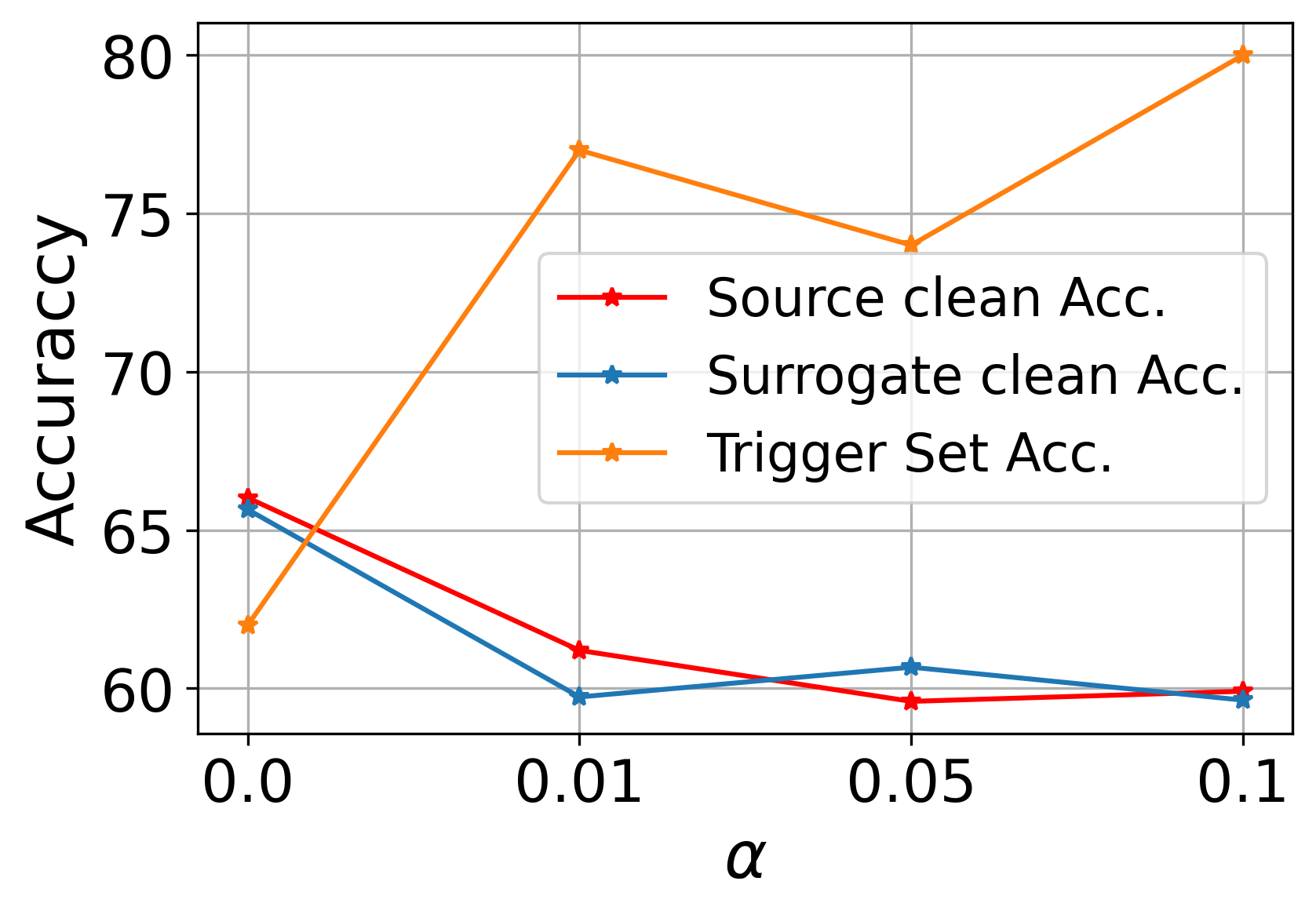}   \\
\end{tabular}
\captionof{figure}{A large $\alpha$ enhances feature regularization, thereby resulting in improved watermarking performance.}
\label{fig: reg_cifar}
\end{minipage}%
\hfill
\begin{minipage}{.48\textwidth}
\begin{tabular}{cc}
  \includegraphics[width=0.5\linewidth]{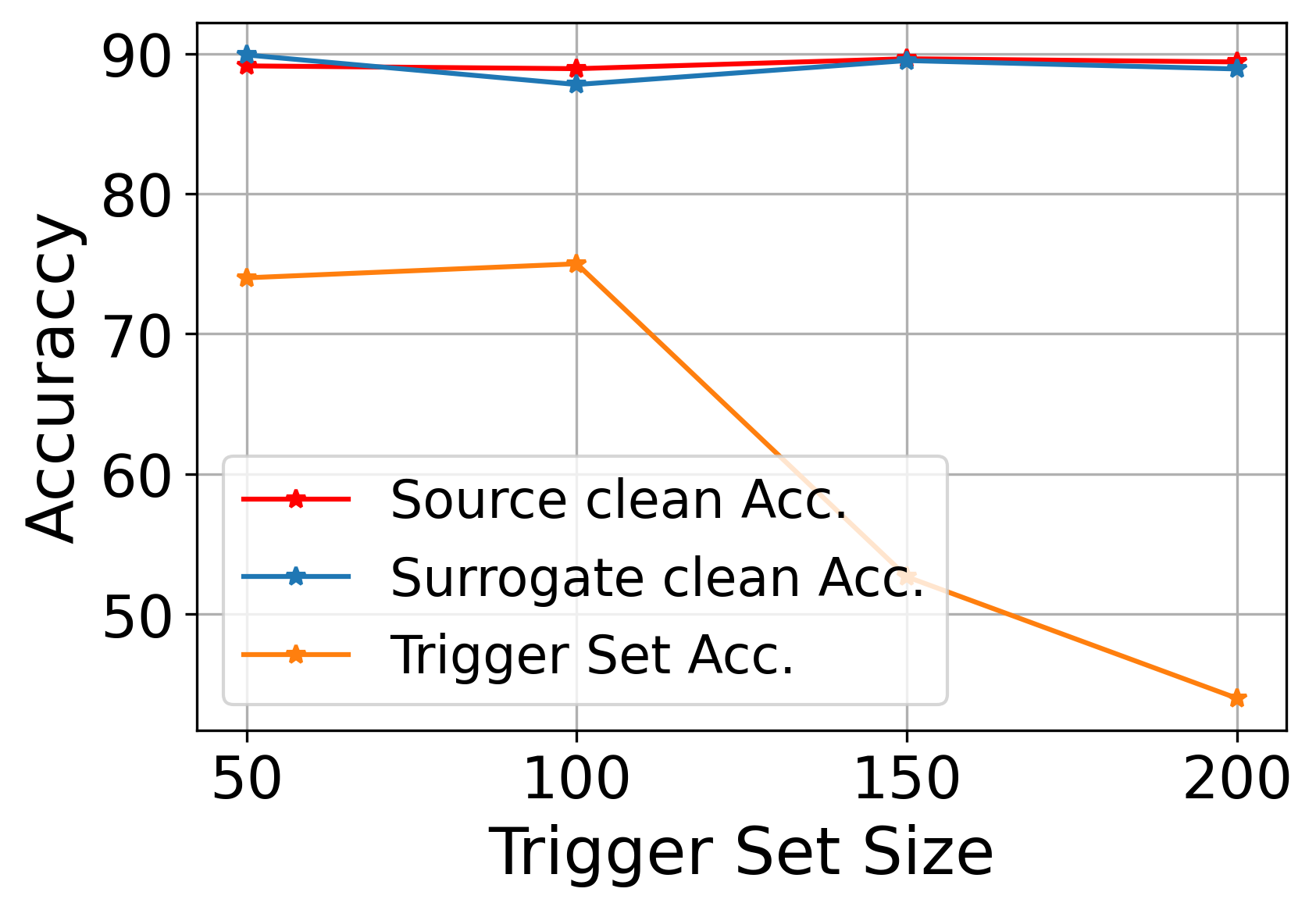}   &  \includegraphics[width=0.5\linewidth]{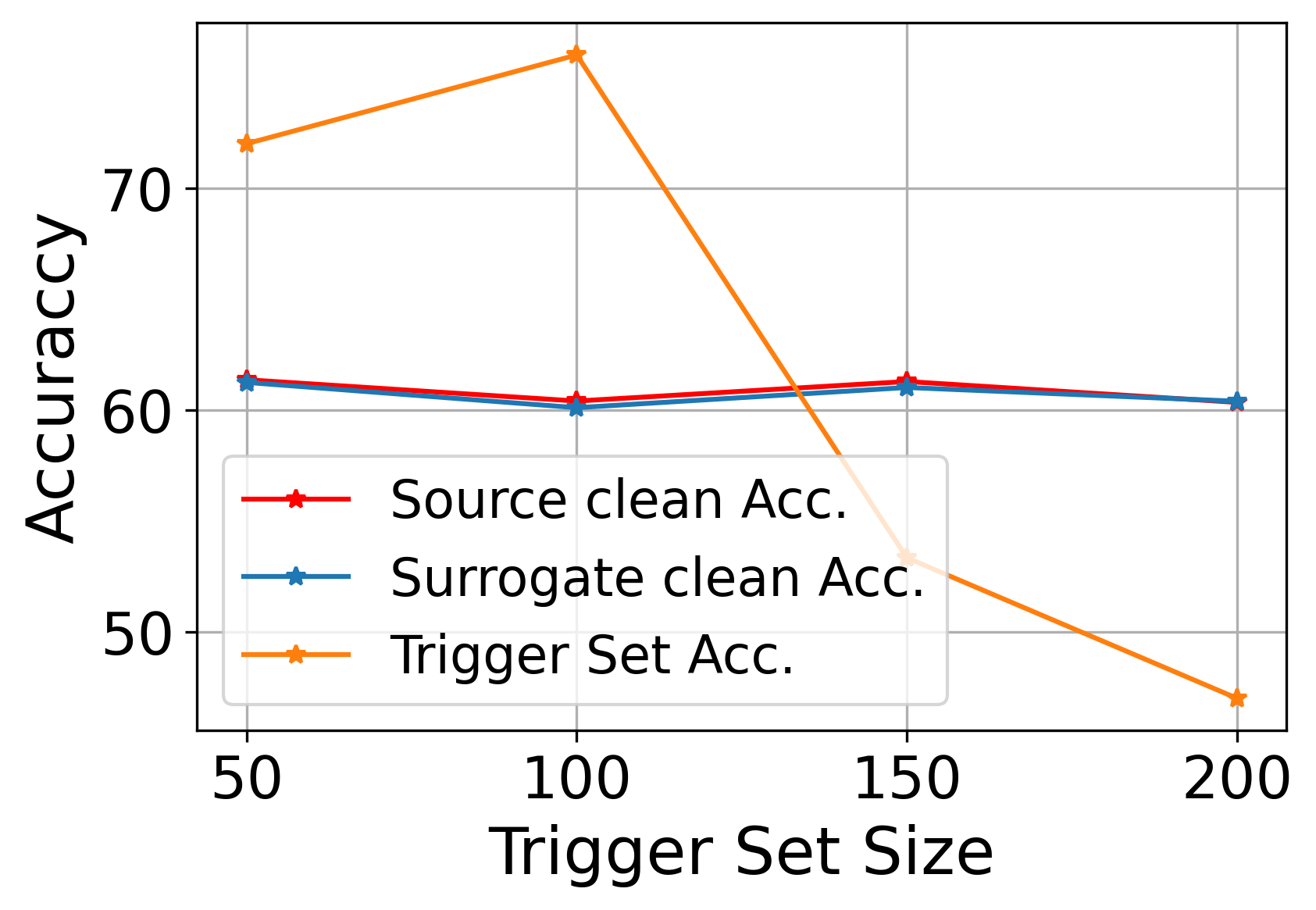}   \\
\end{tabular}
\captionof{figure}{MAT can achieve strong watermarking performance with a small trigger set consisting of multi-view data.}
\label{fig: size_cifar}
\end{minipage}
\end{figure}

\noindent \textbf{Effect of feature regularization.}
%In this section, we provide a detailed analysis of the feature regularization loss employed in Equation \ref{eq: final_loss} during the training of the source model on the trigger set. 
While the inclusion of the feature regularization loss enhances feature learning on the trigger set, it may compromise clean accuracy. In Fig. \ref{fig: reg_cifar}, we explore the impact of varying the balance parameter $\alpha$ within the range of \{0.0, 0.01, 0.05, 0.1\}. The findings indicate that an increase in $\alpha$ does indeed negatively affect clean accuracy. This can be attributed to the fact that features of different classes are not strictly orthogonal, leading to a decrease in clean accuracy. Despite this, an increased value of $\alpha$ significantly enhances the effectiveness of the trigger set for watermarking, aligning with our expectations. In practice, to achieve a balance between clean accuracy and watermarking effectiveness, a smaller value of $\alpha$ is preferred to guide the feature learning process. It is also worth noting that even without the feature regularization loss, MAT still outperforms the margin-based watermarking method which further demonstrates the effectiveness of the trigger set selection strategy.

\vspace{-0.5cm}

\noindent \textbf{Size of the trigger set.} In this section, we vary the size of the trigger set, choosing from \{50, 100, 150, 200\}, to assess the impact of trigger size. Fig. \ref{fig: size_cifar} reveals that varying the trigger set size minimally affects the clean accuracy of both the source model and the surrogate model. However, the accuracy of the trigger set experiences a significant drop as the trigger set size increases. This phenomenon can be easily understood from a feature-learning perspective. Due to the limited instances of samples demonstrating multiple features in the datasets, expanding the size of the trigger set marginally integrates additional multi-view data. Taking CIFAR-10 as an example, when the trigger size is set to 200, there are 88 correctly classified samples within the trigger set. In comparison, for a trigger set size of 100, there are 75 correctly classified samples. Notably, the absolute number of correctly classified samples remains similar.
\vspace{-0.3cm}
\section{Conclusion}
\vspace{-0.5cm}
In this paper, we propose a novel trigger set-based watermarking method, called MAT, using multi-view data. The proposed MAT is easy to interpret and highly effective for watermarking deep neural networks (DNNs). Extensive experiments have been conducted to demonstrate the efficacy of MAT. The findings indicate that MAT not only effectively safeguards against model extraction attacks but also exhibits robustness in the face of white-box attacks, including fine-tuning and fine-pruning. In summary, MAT stands out as a robust watermarking method, offering enhanced protection for the intellectual property of DNNs.

{\bf Acknowledgements.} We would like to thank the anonymous reviewers for their helpful comments. This project was supported by a grant from the University of Texas at Dallas.

\bibliographystyle{splncs04}
\bibliography{main}

\newpage

\clearpage
\section{Appendix}
\vspace{-0.3cm}
In this work, we introduce a novel method for watermarking deep neural networks, based on multi-view data. This appendix to the main paper provides additional elaboration and experimental results on the following:
\begin{itemize}
    \item We illustrate a pipeline figure of our method, MAT, in Section \ref{sec:pipeline}.
    \item We provide more details on various attack methods in Section \ref{sec:details}.  
    \item We include results using distillation attack in Section \ref{sec:distillation}.
    \item We present t-SNE visualizations of the features from the source model and the surrogate models in Section \ref{sec:vis}.
    \item We include the results of other model architectures in Section \ref{sec:dma}.
    \item We explore an alternative strategy for feature regularization in Section \ref{sec: alter}.
    \item We discuss about the multi-view data and feature extraction via an example in Section \ref{sec: multi-view}.
    
\end{itemize}

\subsection{Illustration of MAT}
\label{sec:pipeline}

\vspace{-0.2cm}
 \begin{figure}[htb!]
    \centering
\vspace{-0.4cm}
\includegraphics[width=0.6\textwidth]{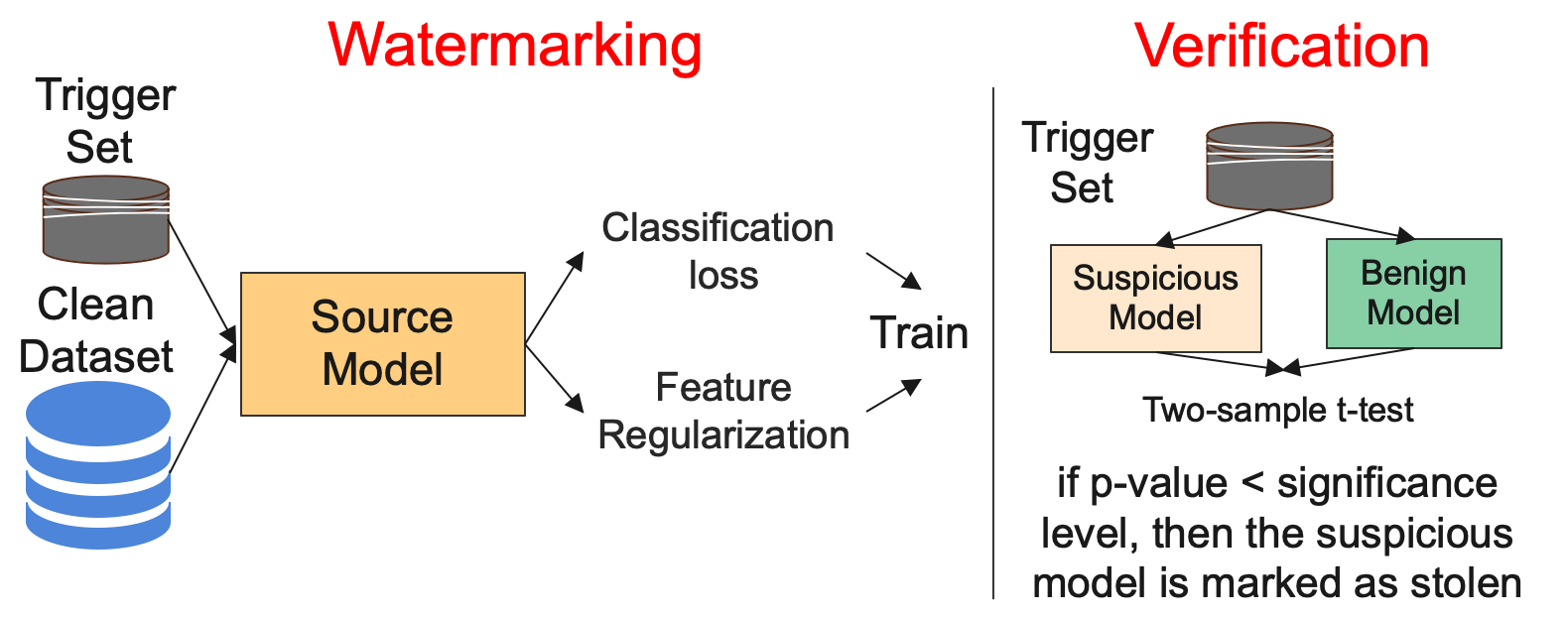}
\vspace{-0.2cm}
    \caption{Pipeline of our proposed method MAT.}
    \label{fig:trigger}
\end{figure}
\vspace{-0.2cm}

Fig. \ref{fig:trigger} illustrates the training pipeline of our proposed method MAT. The details are discussed in the main paper.

\subsection{Implementation Details of Various Attack Methods}
\label{sec:details}
\vspace{-0.1cm}
\begin{itemize}
    \item Model extraction attack \cite{orekondy2019knockoff}:
    When training the surrogate model using the extraction attack, we use the same settings as when training the source model. We use SGD as the optimizer and train 200 epochs with 0.1 as the learning rate and 0.001 as the weight decay.
    \item Fine-tune \cite{liu2018fine}:
    We used the same optimizer as the extraction attack to fine-tune the source model for 100 epochs.
    \item Fine-pruning \cite{liu2018fine}: 
    We perform pruning for the last convolutional layer of the source model and keep subtracting the neurons with the lowest activation on the validation set until the accuracy drops by more than 0.2. Finally, we perform fine-tuning for the pruned model for 100 epochs with the same settings.
\end{itemize}

\subsection{Distillation Attack}
\label{sec:distillation}
\begin{figure}[t]
    \centering
\begin{tabular}{cc}
\multicolumn{2}{c}{\includegraphics[width=0.8\linewidth]{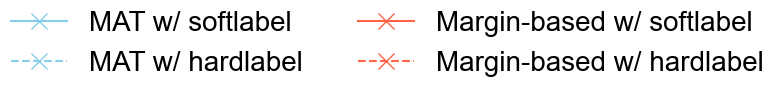}} \\
\includegraphics[width=0.45\linewidth]{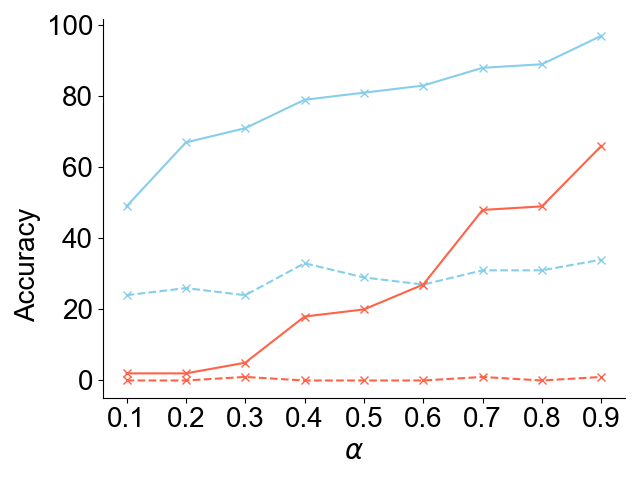}   &     
\includegraphics[width=0.45\linewidth]{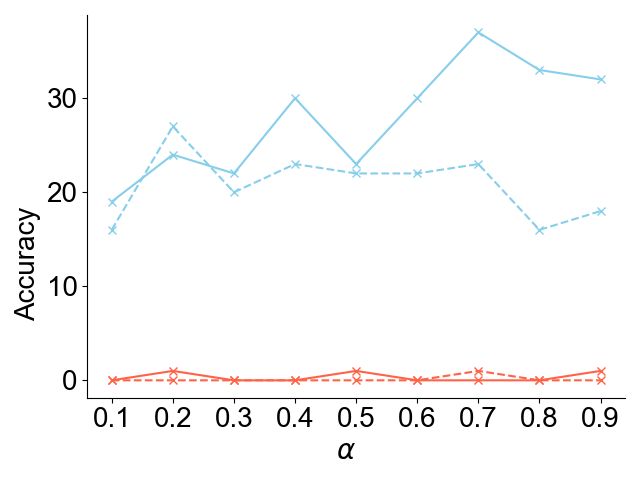} \\
a) CIFAR-10 & b) CIFAR-100
\end{tabular}
\caption{Trigger set accuracy of the surrogate model using soft-label and hard-label model extraction in the distillation attack. MAT has much better watermarking performance compared with the margin-based approach after the distillation attack.}
\label{fig:distill_extraction}
\end{figure}

\vspace{-0.1cm}
Following \cite{kim2023margin}, we also consider a distillation attack that leverages the labels of the surrogate dataset for training the surrogate model. Specifically, the distillation attack proves more powerful than the model extraction attack, as attackers go beyond merely relying on the output of the source model to train the surrogate model. While this makes functionality stealing more challenging, it also renders the process of erasing the watermark more accessible to the attacker. The distillation attack can be formalized as follows,

\begin{equation}
\small
\begin{split}
    \min_{\hat{\theta}} \ell_{\textnormal{distillation}}(\hat{\theta}; \theta, \hat{S} , \alpha) & =  \frac{1}{|\hat{S}|} \sum_{(x,y) \in \hat{S}}  \alpha \cdot D_{KL}(\hat{M}_{\hat{\theta}}(x), M_\theta(x))  \\
    & + (1-\alpha) \cdot  \ell( M_{\hat{\theta}}(x), y)
\end{split}
\label{eq: distill_extraction}
\end{equation}
where $\alpha$ is a hyperparameter that balances the model extraction attack and the standard supervised training on the surrogate dataset. Fig. \ref{fig:distill_extraction} shows the trigger set accuracy of the surrogate model on CIFAR-10 and CIFAR-100 with different choices of $\alpha$. It can be observed that, whether using soft-label or hard-label model extraction in the distillation attack, the proposed MAT outperforms the margin-based approach in terms of watermarking performance. This further underscores MAT's superior ability to defend against distillation attacks when compared to the baseline, highlighting its practical applicability.

\begin{figure*}[t]
    \centering
\begin{tabular}{ccc}
\multicolumn{3}{c}{\includegraphics[width=0.5\linewidth]{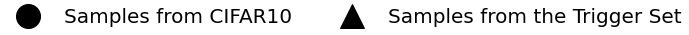}} \\
\includegraphics[width=0.23\linewidth]{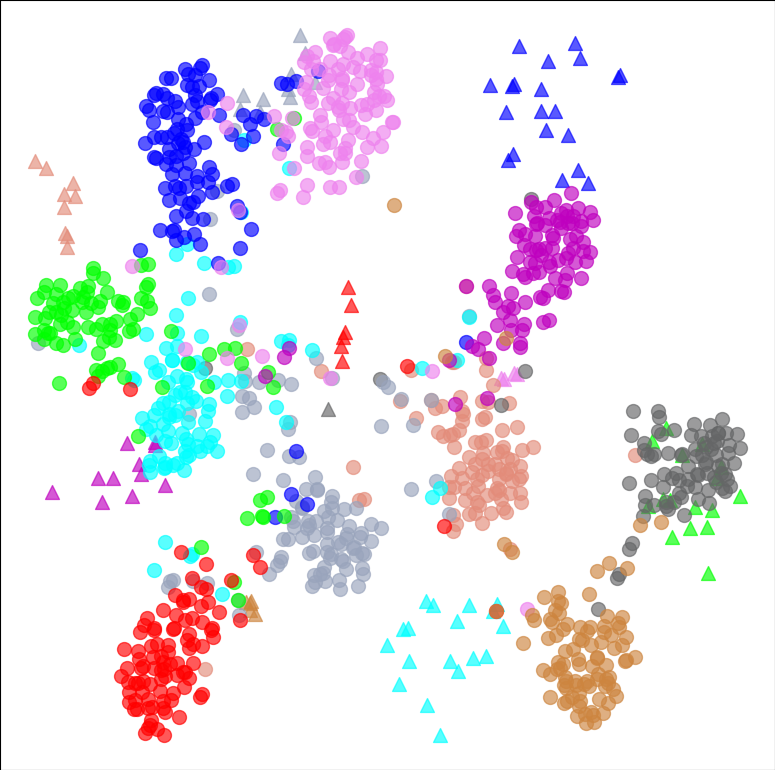}   &  \includegraphics[width=0.23\linewidth]{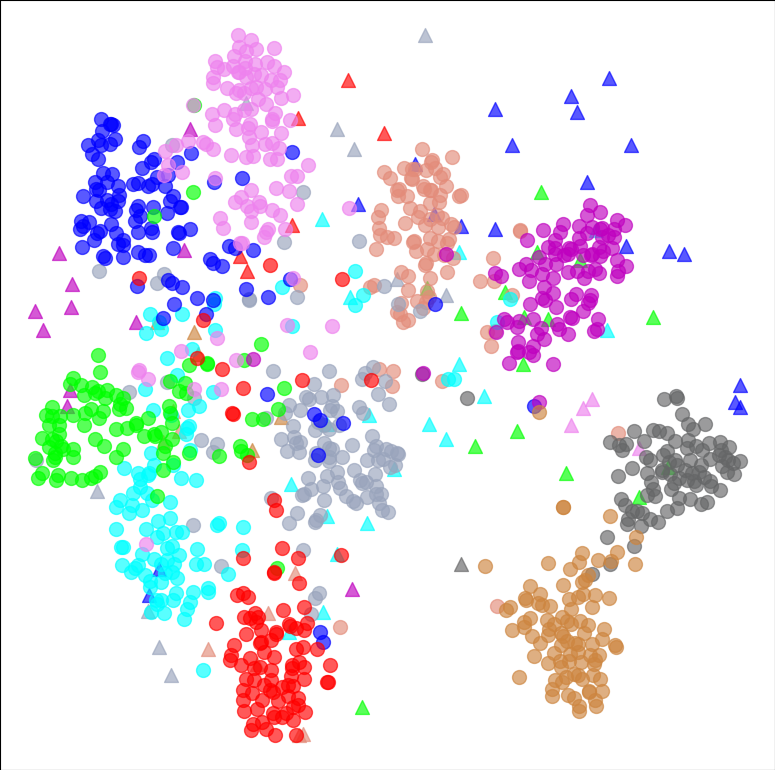}   &   \includegraphics[width=0.23\linewidth]{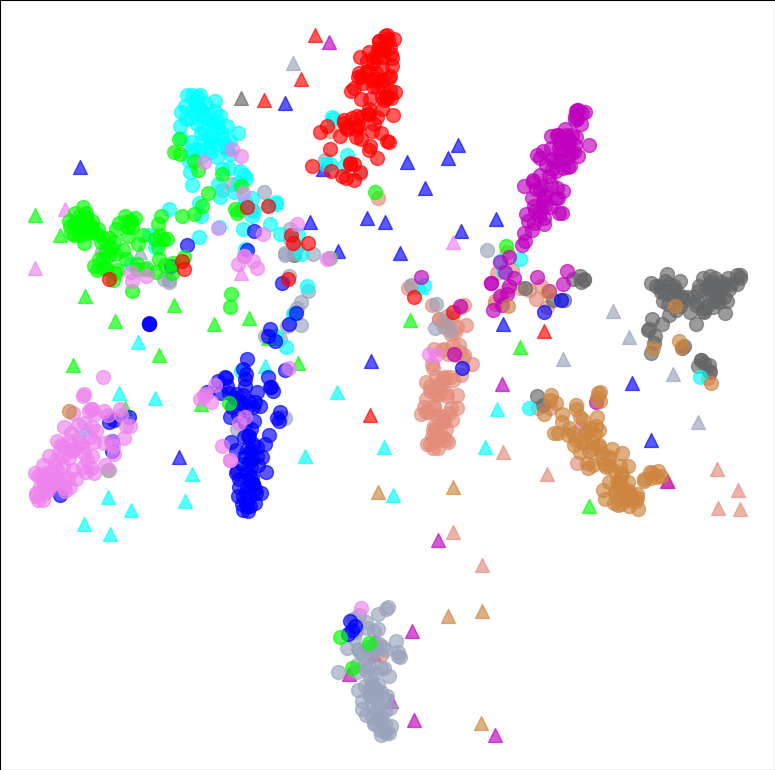}  \\
a) Source& b) Soft-label & c) Hard-label \\
\includegraphics[width=0.23\linewidth]{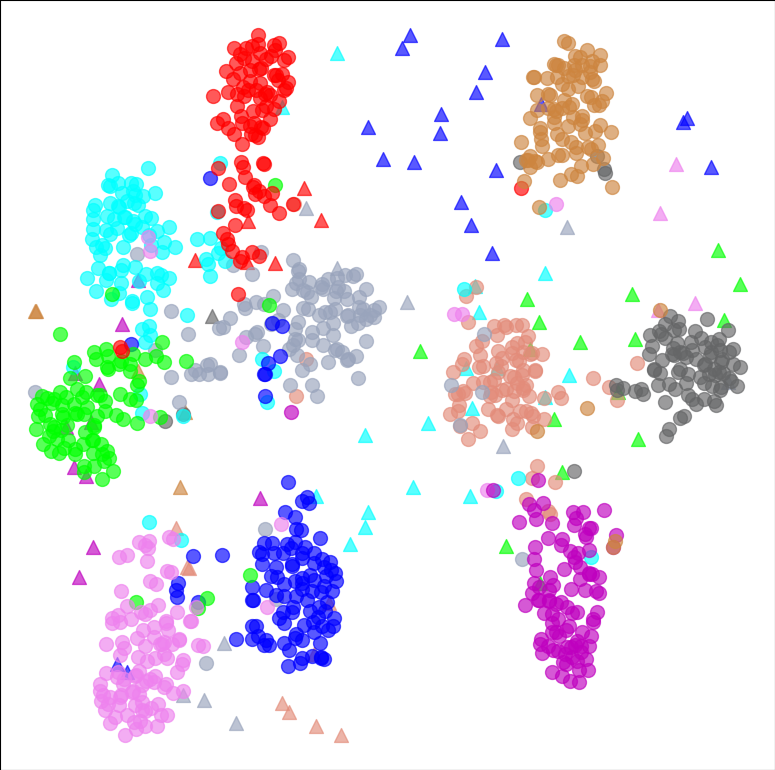}
& \includegraphics[width=0.23\linewidth]{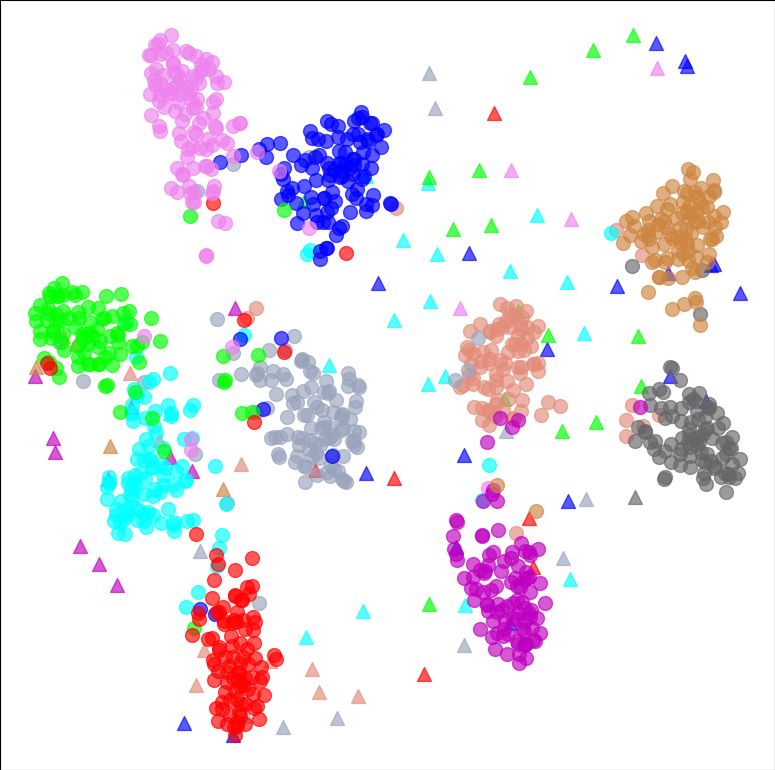}
& \includegraphics[width=0.23\linewidth]{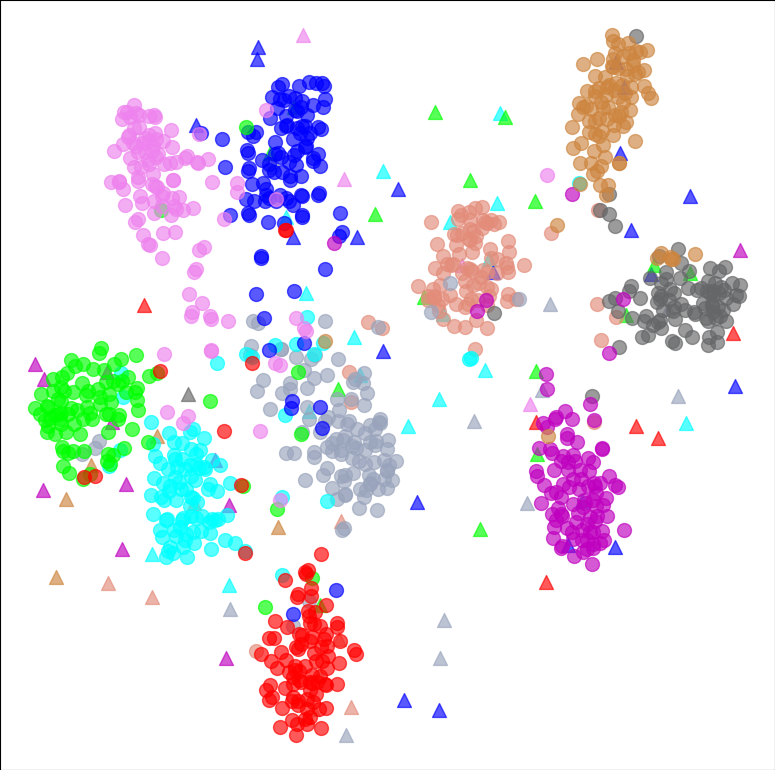}\\
d) Distillation w/ GT & e) Fine-tune & f) Fine-pruning \\
\end{tabular}
\caption{Visualization of the features of the clean samples from CIFAR-10 and the trigger set. The images from the trigger set are close to the decision boundary.}
\label{fig:features}
\end{figure*}

\begin{figure*}[htb!]
    \centering
\begin{tabular}{cc}
\multicolumn{2}{c}{\includegraphics[width=0.5\linewidth]{sec/Figures/legend.png}} \\
\includegraphics[width=0.23\linewidth]{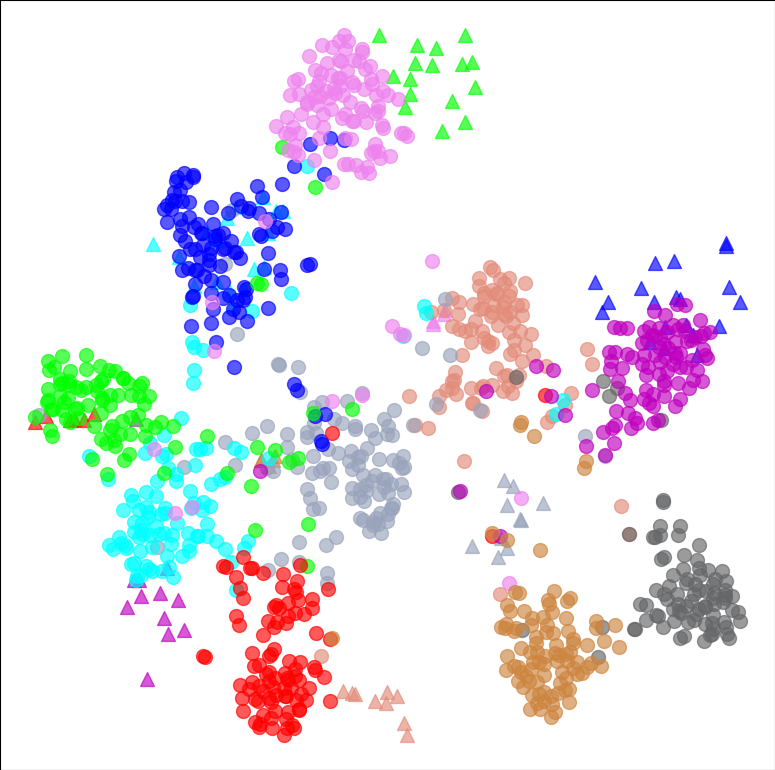}   &  \includegraphics[width=0.23\linewidth]{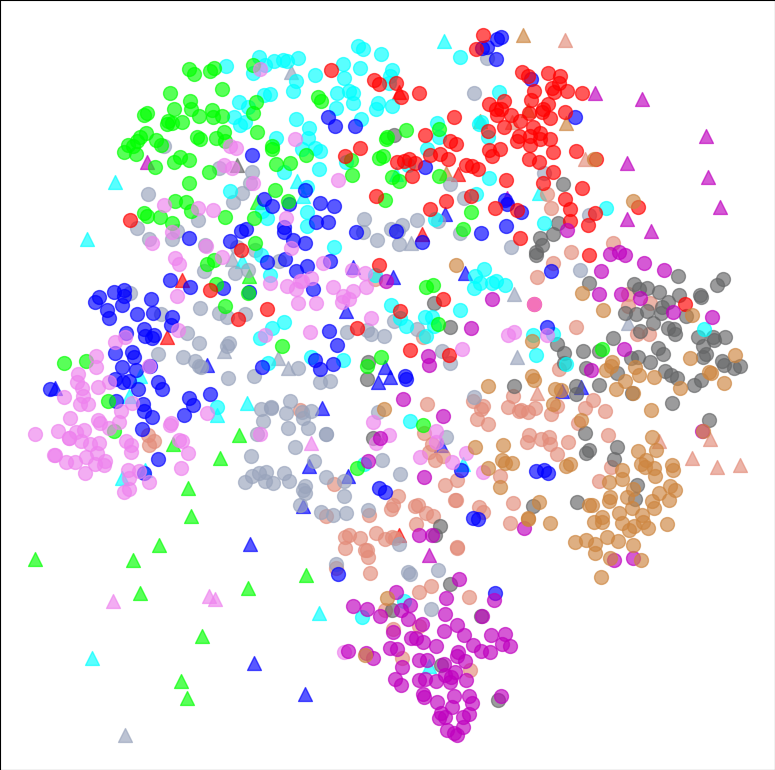}   \\
a) Source& b) Soft-label  \\
\end{tabular}
\caption{Visualization of the features of the clean samples from CIFAR-10 and the trigger set using SVHN as the surrogate dataset. The predictions of the source model on the trigger set can also be transferred to the surrogate model. a) The features of the source model. b) The features of the surrogate model using soft-label functionality attack.}
\label{fig:svhn_features}
\end{figure*}

\subsection{Visualization}
\label{sec:vis}
In this section, we leverage t-SNE \cite{van2008visualizing} to visualize the features of the clean samples from CIFAR-10 and the trigger set. Fig. \ref{fig:features} a) shows the features from the source model trained with the clean training set and the trigger set as in Eq. \ref{eq: source_train}.

% \begin{equation}
%     \ell(M_\theta; S) = \frac{1}{|S|} \sum_{(x,y) \in S} \ell( M_\theta(x), y)
%      \label{eq: source_train}
% \end{equation}
Fig. \ref{fig:features} b), Fig. \ref{fig:features} c), Fig. \ref{fig:features} d), Fig. \ref{fig:features} e) and Fig. \ref{fig:features} f) show the features from the surrogate model using soft-label model extraction attack (Soft-label), hard-label model extraction attack (Hard-label), distillation attack with ground-truth labels (Distillation w/ GT), fine-tune and fine-pruning, respectively. It can be observed that the features of the samples from the trigger set are all close to the decision boundary. Also, the predictions of the source model on the trigger set are transferred to the surrogate model. The visualizations further elucidate the efficacy of the proposed watermarking strategy. Fig. \ref{fig:svhn_features} shows the visualization of the features of the clean samples from CIFAR-10 and the trigger set using SVHN as the surrogate dataset. Similarly, we can observe that the predictions of the source model on the trigger set can also be transferred to the surrogate model.

\begin{table*}[t]
 \caption{Our proposed MAT achieves the best watermarking performance on CIFAR-10 using VGG11 compared with the baselines. }
    \centering
\small
\resizebox{\textwidth}{!}{%
    \begin{tabular}{c|c|c|c|c|c|c|c}
    \toprule

         & \multirow{2}{*}{Source Acc. (\%)}   & \multicolumn{3}{c}{Soft-Label}  & \multicolumn{3}{c}{Hard-Label} \\

         &   & Surrogate Acc. (\%)& Trig. Acc. (\%) & p-value& Surro. Acc. (\%)&  Trig. Acc. (\%) & p-value\\
         \midrule

        Base \cite{zhang2018protecting} &  86.60 &  85.63&  0 &$10^{-1}$& 83.06  &1&$10^{-1}$ \\
            \midrule
        RS \cite{bansal2022certified}  &   88.00& 86.46& 1 &$10^{-1}$& 82.98&3&$10^{-1}$\\
             \midrule
       Margin-based \cite{kim2023margin} &80.20 &  82.03 &16 &$10^{-4}$ & 80.24 &6&$10^{-1}$ \\
    \midrule
        \midrule
    MAT (no feature reg.)  & 86.40& 85.1 & 65 &$10^{-10}$ & 82.01&47  & $10^{-4}$ \\
    \midrule

   MAT & 85.90 & 84.81 & \textbf{68} & $10^{-10}$& 82.06&49&$10^{-4}$\\
    \bottomrule
    \end{tabular}}
    \label{tab:vgg11_cifar10}
\end{table*}

\begin{table*}[t]
\caption{Our proposed MAT achieves the best watermarking performance on CIFAR-10 using ViT-base-patch16-384 \cite{dosovitskiy2020image} compared with the baselines. }
    \centering
\small
\resizebox{\textwidth}{!}{%
    \begin{tabular}{c|c|c|c|c|c|c|c}
    \toprule

         & \multirow{2}{*}{Source Acc. (\%)}   & \multicolumn{3}{c}{Soft-Label}  & \multicolumn{3}{c}{Hard-Label} \\

         &   & Surrogate Acc. (\%)& Trig. Acc. (\%) &p-value& Surro. Acc. (\%)  &Trig. Acc. (\%) & p-value\\
         \midrule

        Base \cite{zhang2018protecting} &  94.90 &  97.34&  0&$10^{-1}$ &  96.72 &0 &$10^{-1}$\\
            \midrule
        RS \cite{bansal2022certified}  &  17.61&22.81 & 10&$10^{-1}$ & 22.74&10&$10^{-1}$\\
             \midrule
       Margin-based \cite{kim2023margin} &92.60 &  96.60 & 1 &$10^{-1}$ & 96.39 & 0&$10^{-1}$\\
    \midrule
        \midrule
    MAT (no feature reg.)  & 97.70 & 97.40 &  55&$10^{-6}$ & 96.84&50&$10^{-4}$ \\
    \midrule

   MAT & 97.60 & 97.56 & \textbf{58} &$10^{-7}$& 96.74&47&$10^{-4}$\\
    \bottomrule
    \end{tabular}}
    \label{tab:vit_cifar10_vit}
\end{table*}
% \vspace{-0.4cm}

\subsection{Different Model Architectures}
\label{sec:dma}
\vspace{-0.1cm}
Tables \ref{tab:vgg11_cifar10} and \ref{tab:vit_cifar10_vit} present the results obtained from employing VGG11 and ViT-base-patch16-384 \cite{dosovitskiy2020image} on CIFAR-10, respectively. It is evident from the comparison that our proposed MAT consistently outperforms other baseline methods, demonstrating its efficacy across diverse model architectures.

\iffalse
\begin{figure*}[htb!]
\centering
\setlength{\tabcolsep}{2pt}
     \includegraphics[width=0.8\linewidth]{sec/Figures/svhn_plot.png}    
    \caption{}
    \label{fig:svhn}
\end{figure*}

\begin{figure*}[t]
\centering
\setlength{\tabcolsep}{2pt}
     \includegraphics[width=0.8\linewidth]{sec/Figures/svhn_plot2.png}    
    \caption{}
    \label{fig:svhn_source}
\end{figure*}
\fi

\subsection{An alternative strategy for feature regularization}
\label{sec: alter}
\vspace{-0.1cm}
In this section, we explore an alternative strategy for feature regularization. In particular, for a given sample $(x_k, y_k,\hat{y}_k)$ in the trigger set, we aim to push the feature of the sample to be away from the mean feature of class $y_k$. This can be achieved by minimizing the following loss,
\begin{equation}
\small
 \min_\theta \ell(\theta; D_{c}) + \ell (\theta; D_{t}) - \alpha \frac{1}{|D_{t}|} \sum_{(x_k,  \hat{y}_k) \in D_{t}} \| f(x_k) - f_{y_k} \|_2. 
 \label{eq: final_loss}
\end{equation}
We use $\alpha = 0.01$ in the experiments. Table \ref{tab:otherstrategy} shows that this strategy achieves similar results to the feature regularization used in the main paper. Essentially, this strategy also encourages the model to learn the features of the modified class to provide better watermarking performance.
 
\begin{table*}[htb!]
\caption{Results using an additional regularization as in Eqn. \ref{eq: final_loss}.}
\centering
\small
\resizebox{0.7\textwidth}{!}{%
    \begin{tabular}{c|c|c|c|c|c}
    \toprule

         & \multirow{2}{*}{Source Acc. (\%)}   & \multicolumn{3}{c}{Soft-Label}   \\

         &   & Surrogate Acc. (\%)& Trig. Acc. (\%) & p-value\\
         \midrule

        ResNet18 on CIFAR-10& 86.90&   86.79& 74&  $10^{-11}$\\
        \midrule
        VGG11 on CIFAR-10&85.90  &84.81  &49 & $10^{-4}$\\
        \midrule
        ViT on ImageNet32&73.32 & 72.50& 71 &$10^{-4}$\\
    \bottomrule
    \end{tabular}}
    \label{tab:otherstrategy}
\end{table*}

\subsection{More detailed discussion and visualization on the multi-view data and feature extraction.} 
\label{sec: multi-view}
\begin{figure}[!h]
\centering
\includegraphics[width=0.6\linewidth]{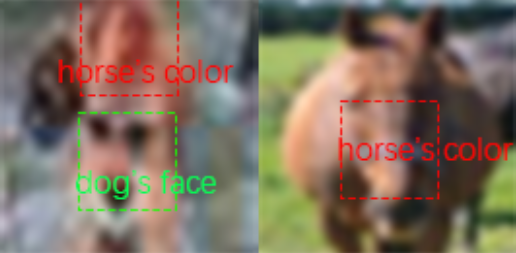}
\caption{\small Dog image with multi-view feature.}
    \label{fig:multi}
\end{figure}

\begin{figure}[!h]
\scriptsize
\centering
\setlength{\tabcolsep}{1pt}
    \begin{tabular}{c c c c}
     \includegraphics[width=0.23\linewidth]{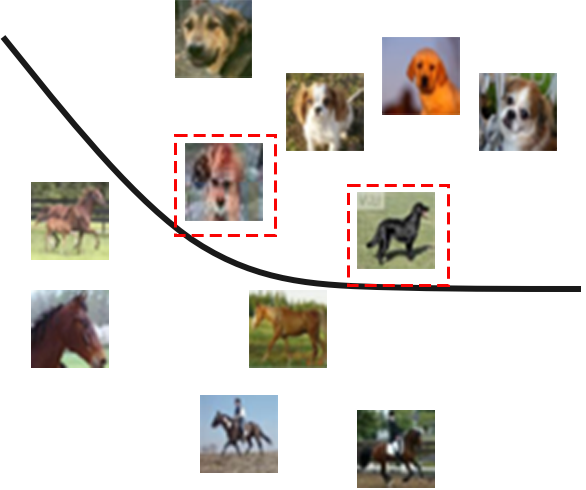}    & \includegraphics[width=0.23\linewidth]{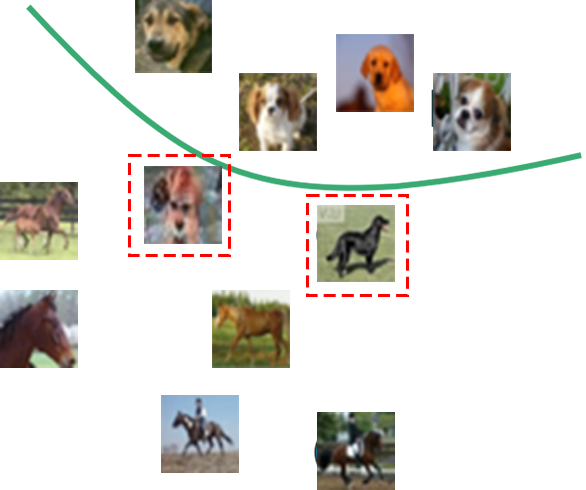} &  \includegraphics[width=0.23\linewidth]{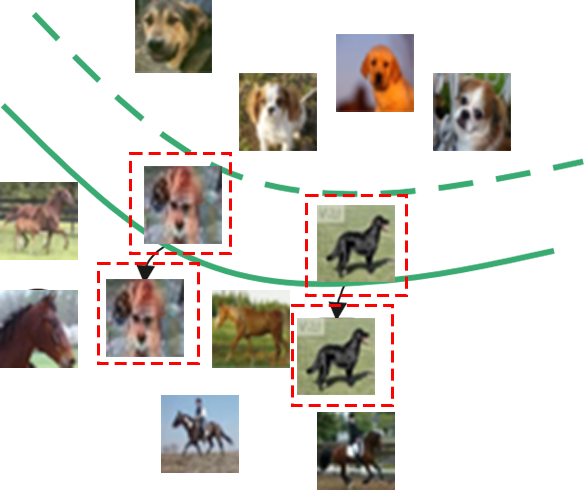}
     & \includegraphics[width=0.23\linewidth]{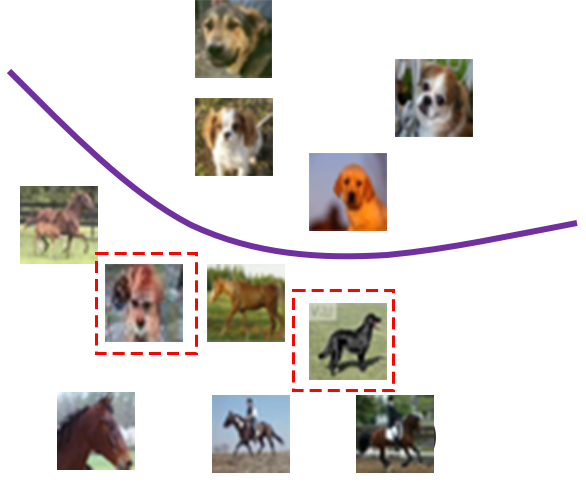} \\
     \scriptsize a)  Source training &\scriptsize  b)  Trigger set training & \scriptsize c) \scriptsize  Feature regularization & \scriptsize d)  \scriptsize 
 Surrogate model \\
    \end{tabular}
\vspace{-0.4cm}
    \caption{ The dog vs. horse example. Multi-view images are shown in the red squares.}
    \label{fig:methods_real_image}
\end{figure}

Figure \ref{fig:multi} shows an example image exhibiting multi-view features. Specifically, the dog image (left-hand side) has a color similar to a horse, which can be leveraged for watermarking the model. Figure \ref{fig:methods_real_image} shows the visualization using the dog vs. horse example to demonstrate the training process.

\vspace{0.3cm}
\vspace{-0.5cm}
% \vspace{-0.4cm}

\end{document}